\documentclass[11pt]{article}

\usepackage[preprint]{acl}

\usepackage{times}
\usepackage{latexsym}
\usepackage{amsmath}

\usepackage[T1]{fontenc}

\usepackage{hyperref}
\usepackage[nameinlink]{cleveref}

\usepackage{xcolor}
\usepackage[table]{xcolor}

\usepackage[utf8]{inputenc}

\usepackage{microtype}

\usepackage{inconsolata}

\usepackage{graphicx}

\usepackage[final]{pdfpages}

\usepackage[most]{tcolorbox}

\usepackage{tabularx}
\usepackage{booktabs}
\usepackage{array}
\usepackage{amssymb}

\usepackage{float}
\usepackage[title]{appendix}
\usepackage{todonotes}
\definecolor{promptbg}{HTML}{ebf5f8}
\definecolor{promptframe}{HTML}{dfeef5}
\definecolor{promptfont}{HTML}{0072b2}

\definecolor{clusterbg}{HTML}{ebf8f0}
\definecolor{clusterframe}{HTML}{dff5ea}
\definecolor{clusterfont}{HTML}{00b238}

\definecolor{safebg}{HTML}{f5ebf8}
\definecolor{safeframe}{HTML}{f4dff5}
\definecolor{safefont}{HTML}{9400b2}

\definecolor{textbg}{HTML}{f8f2eb}
\definecolor{textframe}{HTML}{f5ebdf}
\definecolor{textfont}{HTML}{f78800}

\definecolor{promptbg}{HTML}{ebf5f8}
\definecolor{promptframe}{HTML}{dfeef5}
\definecolor{promptfont}{HTML}{0072b2}

\newcommand{\promptbox}[7]{
    \begin{center}
        \small
        \begin{minipage}{#6\textwidth}
            \begin{tcolorbox}[colback=#1, colframe=#2, arc=4mm, boxrule=0.5pt, leftrule=0pt, rightrule=0pt, toprule=0pt, bottomrule=0pt, width=\textwidth] 
                \textcolor{#3}{#7}
            \end{tcolorbox}
        \captionof{figure}{#4}
        \label{#5}
        \end{minipage}
    \end{center}
}

\usepackage[margin=2.5cm]{geometry}
\usepackage{listings}
\usepackage{xcolor}
\usepackage{booktabs}
\usepackage{array}
\usepackage{tabularx}
\usepackage{minipage-marginpar}
\usepackage{caption}
\usepackage{pgfplots}
\pgfplotsset{compat=1.18}
\usepackage{tikz}
\usepackage[T1]{fontenc}
\usepackage{parskip}
 
\definecolor{codebg}{HTML}{F5F5F2}
\definecolor{commentcolor}{HTML}{6A737D}
\definecolor{keywordcolor}{HTML}{D73A49}
\definecolor{stringcolor}{HTML}{032F62}
\definecolor{funccolor}{HTML}{6F42C1}
\definecolor{humangreen}{HTML}{3B6D11}
\definecolor{modelgray}{HTML}{5F5E5A}
\definecolor{lightgray}{HTML}{EBEBEB}
\definecolor{headerbg}{HTML}{F0EEE8}
\definecolor{bordercolor}{HTML}{CCCCCC}
 
\lstdefinestyle{python}{
  language=Python,
  backgroundcolor=\color{codebg},
  basicstyle=\ttfamily\small,
  commentstyle=\color{commentcolor}\itshape,
  keywordstyle=\color{keywordcolor}\bfseries,
  stringstyle=\color{stringcolor},
  emphstyle=\color{funccolor},
  emph={Fibonacci_Generator},
  showstringspaces=false,
  breaklines=false,
  frame=none,
  xleftmargin=0.75em,
  xrightmargin=0.75em,
  aboveskip=0pt,
  belowskip=0pt,
  columns=flexible,
  keepspaces=true,
  tabsize=4,
}

\usepackage{array}
\usepackage{ragged2e}
\usepackage{booktabs}
\usepackage{arydshln}
\usepackage{makecell}
\usepackage{tabularx}
\usepackage{dblfloatfix}

\usepackage{enumitem}
\setitemize{noitemsep,topsep=0pt,parsep=0pt,partopsep=0pt}
\setenumerate{noitemsep,topsep=0pt,parsep=0pt,partopsep=0pt}

\newcolumntype{L}[1]{>{\RaggedRight\arraybackslash\hspace{0pt}}p{#1}}
\newcolumntype{C}[1]{>{\Centering\arraybackslash\hspace{0pt}}p{#1}}
\newcolumntype{Y}{>{\RaggedRight\arraybackslash\hspace{0pt}}X}


\usepackage{array}
\usepackage{booktabs}
\usepackage{xltabular}
\usepackage{ragged2e}
\usepackage[table]{xcolor}
\usepackage{makecell}

\newcolumntype{Y}{>{\RaggedRight\arraybackslash}X}
\newcolumntype{P}[1]{>{\RaggedRight\arraybackslash}p{#1}}

\renewcommand{\arraystretch}{1.15}

\title{Subjective Code Preferences in Experts and Large Language Models}

\author{
Anna Mokhova,
Subhabrata Dutta,
Iryna Gurevych \and
Simone Balloccu \\
Ubiquitous Knowledge Processing Lab (UKP Lab) \\
Department of Computer Science \\
Technical University of Darmstadt \\
\texttt{anna.mokhova@tu-darmstadt.de}
}

\usepackage{xspace}
\newcommand{\llamasmall}{Llama3-8B\xspace}
\newcommand{\llamabig}{Llama3-70B\xspace}
\newcommand{\qwensmall}{Qwen3-8B\xspace}
\newcommand{\qwenbig}{Qwen3-32B\xspace}
\newcommand{\deepseek}{DS-R1-14B\xspace}
\newcommand{\deepseekcode}{DS-Coder-33B\xspace}
\newcommand{\starcoder}{StarCoder2-15B\xspace}
\newcommand{\qwencoder}{Qwen3-Coder-30B\xspace}
\newcommand{\prometheus}{Prometheus-13B\xspace}
\newcommand{\deepseekgrm}{DS-GRM-16B\xspace}
\newcommand{\gpt}{GPT-5\xspace}
\newcommand{\claude}{Claude-4.5\xspace}
\newcommand{\gemini}{Gemini-3\xspace}

\newcommand{\scorebar}[2]{%
  \begin{tikzpicture}[baseline=0.5ex]
    \fill[lightgray] (0, 0.5ex-0.03cm) rectangle (2.0cm, 0.5ex+0.03cm);
    \foreach \i in {0,1,2,3,4,5}{
      \draw[lightgray!60!black, line width=0.3pt]
        (\i*0.4cm, 0.5ex-0.06cm) -- (\i*0.4cm, 0.5ex+0.06cm);
    }
    \fill[#2] (#1*0.4cm-0.06cm, 0.5ex-0.09cm)
              rectangle (#1*0.4cm+0.06cm, 0.5ex+0.09cm);
  \end{tikzpicture}%
}

\setlist[enumerate]{leftmargin=*, itemsep=0.5em, topsep=0.6em}

\begin{document}

\maketitle
\begin{abstract}

Large Language Models (LLMs) have become increasingly popular for coding tasks, with subjective coding preferences being an essential element to adapt to programmers’ personal needs. Existing work overlooks such characteristics and mainly focuses on code correctness. In this study, we propose a typification of four subjective coding preference axes — complexity, commenting, modularity, and readability — motivated by common engineering habits and validated by 25 software engineers. We collect a dataset of $\approx3,000$ paired Python code snippets reflecting these axes, annotated by 73 experts who rate their preferences on a Likert scale. Using our dataset, we study how LLMs handle subjective coding preferences. We present 13 LLMs with pairs of solutions to the same programming task, first as textual descriptions and then as concrete code snippets. We find that models often prefer one option in natural language but the opposite when evaluating code. More consistent models (i.e., those that are coherent in their choices between deeds and words) frequently reveal positional bias: swapping the order of options changes the preferred alternative. We then use the five most consistent models to re-annotate the dataset. Compared to humans, models show polarized Likert distributions and notable divergence in ratings. 
A case study on \gpt reveals reliance on external assumptions and brittle reasoning.\footnote{Data available at \href{https://anonymous.4open.science/r/CodePref-B07D/}{https://CodePref-B07D}}
\end{abstract}

\begin{figure*}[t]
\centering
 \includegraphics[width=\linewidth]{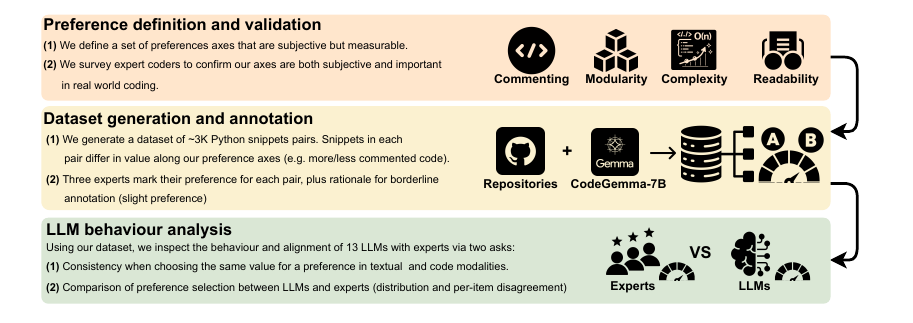}
  \caption{Overview of our dataset construction and experimental pipeline.}
  \label{fig:overview}
\end{figure*}

\section{Introduction}

In recent years, LLMs have become broadly used as coding tools \cite{ferino2025novice}. To measure their coding capabilities, there is a need to assess the models' prior knowledge and skills. This is required to differentiate between the system's fundamental understanding of the task and its generalization power \cite{chollet2019measure}, which might be specious at its logical core.

LLM evaluation in the coding domain typically involves benchmarks that target objective aspects such as code correctness and security \cite{weyssow2026codeultrafeedback, liu2024learning}. This, however, limits the estimation to a "correct vs incorrect" paradigm and does not explain how LLMs behave when subjective coding preferences come into play. This is especially critical since LLMs can be inconsistent in opinion-based or value-related domains \cite{xu2025large}. Focusing only on objective criteria also contradicts the fundamental definition of a \textit{preference} \cite{brafman2009preference}, which is a set of \textit{alternative} choices, where a more desirable choice precedes a less desirable one. Finally, evaluating code quality does not directly validate the model's core understanding of the task that it is performing \cite{moskvichev2023conceptarc}.


In this work, we provide an extensive investigation of LLMs' behavior across subjective coding preferences (Figure \ref{fig:overview}). First, we introduce a set of coding preferences axes that are subjective (i.e., both preference values produce a working code) but can still be measured and shown from the previous work to be important for software development \cite{tashtoush2023notional, de2005study}: commenting, modularity, readability, and complexity. 
We then generate a novel dataset of $\approx3000$ programming problems with paired solution snippets, where each snippet of a pair represents a different value on our preference axes (e.g., more vs less commented code to solve the same problem). This dataset is then fully annotated by 73 coding experts, using a Likert-6 scale to express their subjective preferences for each pair. Borderline annotations (slight preferences) are enriched by experts with textual rationale.

We use our generated dataset to analyze the behavior and alignment of 13 LLMs in two setups. First, we test if LLMs are consistent in expressing a utility at a textual level (e.g., "a more/less commented code") and code level (coding snippet solving the problem with more/less comments). All tested LLMs show severe inconsistency, choosing a specific textual preference, but then picking the opposite coding snippet, and most of the evaluated models exhibit positional bias. 

Second, we let LLMs replicate our dataset annotation, and compare their preference distribution with that of experts: most models 
predominantly choose the extremes of our selection scale), compared to the more balanced one of our experts. \gpt, the only exception, more closely models the overall human distribution, but still shows severe divergence on a per-item basis. We also run a case study where we qualitatively inspect cases of disagreement between experts and \gpt, showing reliance on external assumptions 
and brittle reasoning, culminating in ignoring core knowledge necessary for a correct assessment.

To summarize, this paper provides the following three contributions (Figure \ref{fig:overview}): 

\begin{enumerate}[leftmargin=*, itemsep=0.25em, topsep=0.25em, parsep=0pt]
    \item A novel dataset of Python coding snippets showcasing subjective preferences, validated and annotated by expert software engineers (SE) (\S\ref{sec:data}).
    \item First assessment of LLM consistency over subjective coding preferences, across natural language and code modality (\S\ref{sec:eval_llm_consistency}).
    \item Analysis of the alignment and behavior between LLMs and humans with respect to subjective coding preferences (\S\ref{sec:preference_behavior_llms_humans}).
\end{enumerate}

\section{Related Work}
\paragraph{Coding Preferences}
Previous works related to LLMs’ coding preferences \cite{weyssow2026codeultrafeedback, liu2024learning} inspect objective criteria, such as security or correctness
\cite{liu2024learning, weyssow2026codeultrafeedback}. These studies do not reflect the complexity of decisions and tradeoffs that a developer experiences \cite{dutta2025problem}.
This tendency is also mirrored in existing datasets like CodeUltraFeedback \cite{weyssow2026codeultrafeedback}, CodePrefBench \cite{liu2024learning}, and OpenCodeInstruct \cite{ahmad2025opencodeinstruct}, which focus on tuning LLMs for software development problems, but not on preference alignment. It results in LLMs developing towards code generation correctness and ignoring humans' priority for a more nuanced and complex approach to software development \cite{azeem2025ai}. Some examples of the importance of this multi-criteria analysis include readability's enhancement of software maintainability \cite{buse2009learning, tashtoush2023notional}, or development optimization across multiple objectives \cite{li2024methodology}. While some previous research integrates benchmarking to assess coding criteria beyond correctness \cite{zheng2024beyond}; it does not provide validation from the side of the end users, as opposed to our study.


\paragraph{Human-LLM preferences comparison}
Being trained on vast amounts of natural language data, LLMs show similar decision heuristics to humans in text generation \cite{suri2024large}, including preexisting biases and conceptual behavior tendencies. However, this human-like pattern is not exhibited in controlled binary-choice experiments, where the decision-making requires broader narrative support \cite{jia2024decision} and is systematically distinguished by positional bias \cite{shi2024judging}. This leads to LLMs being more prone to hesitating and admitting their limits in the presence of a clear dislike \cite{li2024dissecting}. Additionally, AI systems are argued to better maximize the satisfaction of preferences and uncover hidden contexts by integrating normative reasoning with machine learning \cite{zhi2025beyond}. In this experiment, we enrich borderline preferences with textual rationale to gain a better, first-of-its-kind understanding of how humans and LLMs ground their choices with regard to decision-making in the coding domain.

\paragraph{LLM Alignment on Code and Natural Language}
Prompt-based inference \cite{song2023comprehensive, wei2022chain} has been shown to enhance LLMs' performance across a variety of coding tasks, mostly generation \cite{jain2022jigsaw,weyssow2025exploring} and testing \cite{kang2023large, lemieux2023codamosa}. 
LLMs also systematically fail in correctly performing “description-to-code” evaluations due to inherent biases and hallucinations within LLMs \cite{jin2025uncovering}. There is no up-to-date proof of the positive contributions of these methodologies to LLMs' alignment between words and coding deeds. We address this in our study by exploring different prompting setups.

\section{Data}\label{sec:data}

\subsection{Coding Preferences}

We define four coding preference axes, designed to be subjective but measurable. This means that choosing one value of the axis does not affect the code correctness; instead, the choice for each preference should be motivated by the individual professional experience, project requirements, or specific industry protocols. For our axes, we model:

\paragraph{\textbf{Commenting}} Appropriate code commenting benefits code maintenance, and speeds up debugging. However, excessive or redundant comments can be detrimental. \textit{'Commenting'} utility represents the extent to which a code snippet contains commented lines, including the code documentation \cite{rani2023decade, perera2023analyzing}.

Consider a code snippet $\mathcal{S}$, with 
$\mathcal{L}_{\mathrm{total}}(\mathcal{S})$ total lines, 
and $\mathcal{L}_{\mathrm{comment}}(\mathcal{S}) \subseteq \mathcal{L}_{\mathrm{total}}(\mathcal{S})$ 
commented lines. The commenting score $CS(\mathcal{S})$ is:
\begingroup\small
\[
\mathrm{CS}(\mathcal{S})
=
\frac{\left|\mathcal{L}_{\mathrm{comment}}(\mathcal{S})\right|}
     {\left|\mathcal{L}_{\mathrm{total}}(\mathcal{S})\right|}.
\]
\endgroup

\paragraph{\textbf{Modularity}} Code modularity refers to the practice of breaking down a large program into smaller, independent, and reusable units called modules. Making a correct and balanced use of this coding feature can improve code organization, re-usability, and maintainability. Modularity is typically measured by the number of separate modules and functions, divided by the total number of code lines \cite{hughes1989functional, tempero2023experiment}. Specifically, given a snippet $\mathcal{S}$, with 
$\mathcal{M}(\mathcal{S})$ distinct modules and functions, and
$\mathcal{L}_{\mathrm{total}}(\mathcal{S})$ lines of code, the modularity score of $\mathrm{MS}(\mathcal{S})$ is:

\begingroup\small
\[
\mathrm{MS}(\mathcal{S})
=
\frac{\left|\mathcal{M}(\mathcal{S})\right|}
     {\left|\mathcal{L}_{\mathrm{total}}(\mathcal{S})\right|}.
\]
\endgroup

\paragraph{\textbf{Complexity}} The degree of complexity of a code can have a profound effect on maintainability \cite{tashtoush2023notional}, developer productivity, software quality, performance, and overall business cost. For this preference, we consider coding snippets with opposite complexity codes in terms of big O-notation \cite{sikka2020learning}. Consider a program $\mathcal{P}$ with input size $n$, 
$T_{\mathcal{P}}(n)$ time complexity, and $S_{\mathcal{P}}(n)$ space 
complexity functions. The asymptotic complexity class of $\mathcal{P}$ is characterized by:

\begingroup\small
\[
\mathcal{C}(\mathcal{P})
=
\left(
\mathcal{O}\!\left(T_{\mathcal{P}}(n)\right),
\mathcal{O}\!\left(S_{\mathcal{P}}(n)\right)
\right),
\]
\endgroup

\paragraph{\textbf{Readability}} Code readability is a fundamental coding characteristic \cite{borstler2023developers, wyrich202340} that defines developers’ preferences and values in structuring their code. However, it is very subjective and experience-based, without an established unified measurement \cite{scalabrino2017automatically}. One definition of readability is linked to judging how easy a text is to understand \cite{buse2009learning}, which can be represented in overt code features (e.g., average or maximal line length, number of loops, etc.). On the other hand, discussions on this topic \cite{rosenberg1998software, chidamber1994metrics, oliveira2024understanding} agree that following the coding standards, cyclomatic complexity, and code maintainability are also associated with the concept of readability. Information on our proposed Readability metric, which combines above-mentioned factors, can be found in the Figure \ref{fig:readability} and in the Appendix \ref{section:readability_metric}.


\begin{figure*}[t]
\centering
\includegraphics[width=\linewidth]{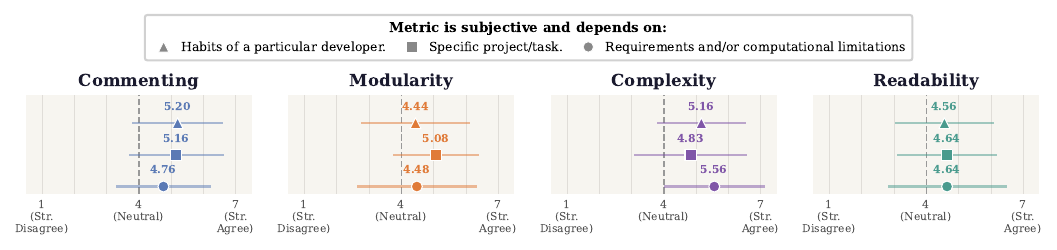}
  \caption {Responses from the experts' survey for each preference: aggregated values, along with context-specific importance (project, requirements, or habits). Appendix \ref{app:survey} contains the full questionnaire and Likert distribution.}
  \label{fig:survey}
\end{figure*}

\paragraph{\textbf{Axes validation}}
The goal of our work is to investigate preferences that matter in real-world coding. Therefore, we validate our four axes through an online survey with experts (recruitment details in Appendix \ref{app:ethics_statement}; full survey in Appendix \ref{app:survey}).
We recruit 25 coders from Upwork\footnote{\href{https://www.upwork.com/}{upwork.com}} with heterogeneous professional background (tab:participant-background). We ask the experts to rate, on a Likert-7 scale, the importance of our four axes and their dependency on personal coding habits, project constraints, and requirements. From results (\Cref{fig:survey}), all metrics are confirmed as influenced by company/project requirements, and experts show average agreement on their importance in different software engineering environments. 

\subsection{Generating the Dataset}
Next, we realize our preferences axes as a dataset $\mathcal{D}$ of paired code snippets

\[
\mathcal{D}
=
\left\{
\left(p, (s_1, d_1), (s_2, d_2)\right)
\right\},
\]

where $p$ is a programming problem description, 
$s_i$ is a code snippet that solves $p$, and 
$d_i$ is a structured description of $s_i$ with respect to a specific evaluation metric ("e.g., minimal VS extensive commenting"). Each pair $(s_1, s_2)$ represents alternative solutions to the same problem $p$, 
annotated by their corresponding metric-based characterizations $(d_1, d_2)$.

\begin{figure*}[ht]
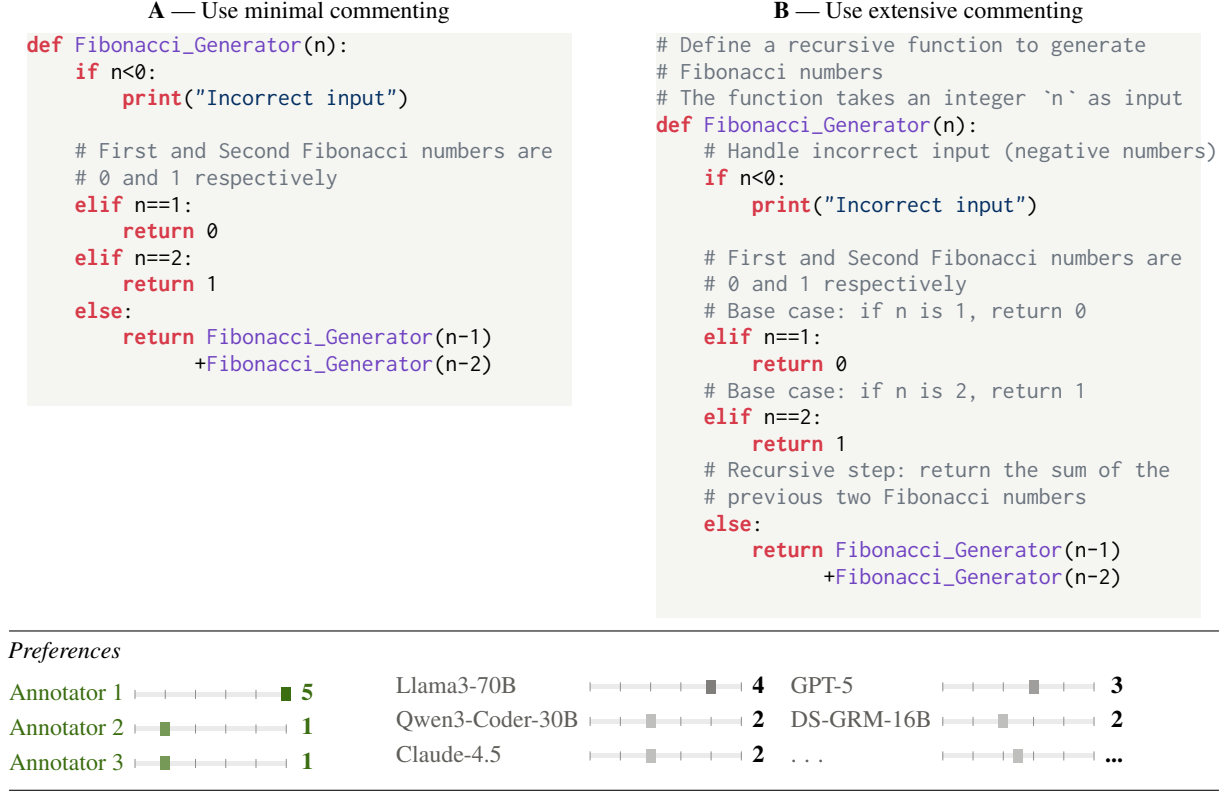

\centering\small
\noindent
\begin{minipage}[t]{0.48\linewidth}
  \centering\textbf{A} — Use minimal commenting\\[3pt]
  \lstset{style=python, basicstyle=\small\ttfamily}
  \begin{lstlisting}
def Fibonacci_Generator(n): 
    if n<0: 
        print("Incorrect input") 
  
    # First and Second Fibonacci numbers are
    # 0 and 1 respectively
    elif n==1: 
        return 0
    elif n==2: 
        return 1
    else: 
        return Fibonacci_Generator(n-1)
              +Fibonacci_Generator(n-2)
  \end{lstlisting}
\end{minipage}%
\hfill
\begin{minipage}[t]{0.48\linewidth}
  \centering\textbf{B} — Use extensive commenting\\[3pt]
  \lstset{style=python, basicstyle=\small\ttfamily}
  \begin{lstlisting}
# Define a recursive function to generate 
# Fibonacci numbers
# The function takes an integer `n` as input
def Fibonacci_Generator(n): 
    # Handle incorrect input (negative numbers)
    if n<0: 
        print("Incorrect input") 
  
    # First and Second Fibonacci numbers are 
    # 0 and 1 respectively 
    # Base case: if n is 1, return 0
    elif n==1: 
        return 0
    # Base case: if n is 2, return 1
    elif n==2: 
        return 1
    # Recursive step: return the sum of the 
    # previous two Fibonacci numbers
    else: 
        return Fibonacci_Generator(n-1)
              +Fibonacci_Generator(n-2)
  \end{lstlisting}
\end{minipage}

\vspace{0.4em}
\noindent\rule{\linewidth}{0.4pt}

\begin{minipage}[t]{0.28\linewidth}
  \textit{{Preferences}}\\[4pt]
  \renewcommand{\arraystretch}{1.3}
  \begin{tabular}{@{}l@{\hspace{4pt}}l@{\hspace{4pt}}r@{}}
    \textcolor{humangreen}{Annotator 1} & \raisebox{0.8mm}{\scorebar{5}{humangreen}}    & \raisebox{0.1mm}{\textbf{\textcolor{humangreen}{5}}} \\
    \textcolor{humangreen}{Annotator 2} & \raisebox{0.8mm}{\scorebar{1}{humangreen!70}} & \raisebox{0.1mm}{\textbf{\textcolor{humangreen}{1}}} \\
    \textcolor{humangreen}{Annotator 3} & \raisebox{0.8mm}{\scorebar{1}{humangreen!70}} & \raisebox{0.1mm}{\textbf{\textcolor{humangreen}{1}}} \\
  \end{tabular}
\end{minipage}%
\hfill
\begin{minipage}[t]{0.68\linewidth}
  \vspace{4pt}
  \renewcommand{\arraystretch}{1.3}
  \begin{tabular}{@{}l@{\hspace{4pt}}l@{\hspace{4pt}}r@{\hspace{10pt}}l@{\hspace{4pt}}l@{\hspace{4pt}}r@{}}
        \textcolor{modelgray}{\llamabig}      & \raisebox{0.8mm}{\scorebar{4}{modelgray!80}} & \textbf{4} &
    \textcolor{modelgray}{\gpt}        & \raisebox{0.8mm}{\scorebar{3}{modelgray!60}} & \textbf{3} \\
    \textcolor{modelgray}{\qwencoder}  & \raisebox{0.8mm}{\scorebar{2}{modelgray!40}} & \textbf{2} &
    \textcolor{modelgray}{\deepseekgrm} & \raisebox{0.8mm}{\scorebar{2}{modelgray!40}} & \textbf{2} \\
    \textcolor{modelgray}{\claude}           & \raisebox{0.8mm}{\scorebar{2}{modelgray!40}} & \textbf{2} & \textcolor{modelgray}{. . .}           & \raisebox{0.8mm}{\scorebar{2.5}{modelgray!40}} & \textbf{...}\\
  \end{tabular}
\end{minipage}

\vspace{0.3em}
\noindent\rule{\linewidth}{0.4pt}
\caption{Examples of an instance from our dataset. Given a Python problem (e.g., generating the Fibonacci numbers), we provide (1) two solution snippets varying across a preference axis (e.g., more or less commented); (2) preference annotation from human experts and LLMs.}
\label{fig:dataset}
\end{figure*}

\begin{table}[t]
\centering
\small
\caption{Participant background and coding context.}
\label{tab:participant-background}
\begin{tabular}{lr}
\toprule
\textbf{Category} & \textbf{Share} \\
\midrule
\multicolumn{2}{l}{\textbf{Development background}} \\
Backend development & 28\% \\
Frontend development & 24\% \\
Full-stack development & 20\% \\
Machine learning engineering & 16\% \\
Other fields & 12\% \\
\midrule
\multicolumn{2}{l}{\textbf{Coding context}} \\
Company projects & 88\% \\
Academic or freelance development & 12\% \\
\bottomrule
\end{tabular}
\end{table}

\begin{table}[ht]
  \centering
  \small
  \caption{Dataset composition by preference axis.}
  \label{tab:dataset}
  \begin{tabular}{lrr}
    \toprule
    \textbf{Axis} & \textbf{Portion} & \textbf{Snippets pairs} \\
    \midrule
    Commenting   & 26\% &  765 \\
    Complexity   & 35\% & 1{,}030 \\
    Modularity   & 24\% &  706 \\
    Readability  & 15\% &  442 \\
    \midrule
    \textbf{Total} & \textbf{100\%} & \textbf{2{,}944} \\
    \bottomrule
  \end{tabular}
\end{table}

For Commenting and Modularity, we first extract source code snippets from existing Python programming datasets \cite{cassano2023can, muennighoff2024octopack, tarun2021pythoncodeinstructions18k, kaggle2023leetcode}. We select snippets that score low on our metrics (e.g, for \textit{Modularity}, we select only those samples that do not contain any modules, thus with $\mathrm{MS}(\mathcal{S})=0$). Then, we prompt CodeGemma-7B \cite{team2024codegemma} to generate the opposite of the given snippet (as in, high-scoring), and manually review the resulting code (prompt in \Cref{figure:prompts_codegemma} in Appendix \ref{sec:prompts}). We manually review the generated snippets to make sure they do not stylistically diverge from the original ones (more details in Appendix \ref{section:codegemma7b}). 

For Complexity and Readability, coding pairs are acquired from open-access coding competitions repositories \cite{thealgorithms_github, gfg_dsa_a2z} and are fully human-written; thus, no LLM is used for code generation. The final dataset contains 2944 coding pairs (statistics in \Cref{tab:dataset}). We also conduct statistical testing, verifying that the produced coding snippets score significantly and consistently different on our metrics (Appendix \ref{section:validating_variety}). 

\subsection{Human Preference Annotation}\label{sec:human_preference_annotation}
To compare humans and LLMs, we collect human preferences annotations for our dataset. We hire experts from Prolific\footnote{\href{https://www.prolific.com/}{prolific.com}}, with a background in software engineering and a degree in computer science. We verify their ability to differentiate preferences via a qualification task: given 8 coding pairs, experts are asked to mark those with a higher/lower degree of commenting/modularity or are more/less computationally complex. Only annotators with 100\% passing rate are accepted.

\paragraph{\textbf{Preference annotation}}
In the main task (example of the annotation UI in Appendix \ref{sec:annotationweb}, \Cref{fig:annotation}), annotators are shown (1) a textual description of a Python problem; (2) two solutions (code snippets) solving it; (3) a clarification of the differences between them at the preference axis level (e.g. "Snippet A/B uses minimal/extensive commenting"); (4) a brief explanation of the axis (e.g. "Commenting refers to the commented of docstring lines..."). Experts choose their preference among the two snippets, based on their working experience, via a Likert-6 scale (0 = Strongly prefer snippet A; 5 = Strongly prefer snippet B). The median option was deliberately omitted to obtain distinguishable preferences between A and B, since, otherwise, humans lean towards the neutral option \cite{kusmaryono2022number}, while LLMs incline towards hesitating \cite{li2024dissecting}. Final numeric preference annotation is done by 73 annotators, where 27 experts annotated $\approx$176 coding pairs each, and 46 annotated $\approx$89 coding pairs each. Each coding pair is annotated 3 times by different experts, and code pairs were presented in a randomized order. \Cref{fig:dataset} shows an instance example (including LLMs preferences, which we cover in \S\ref{sec:preference_behavior_llms_humans}).

\paragraph{\textbf{Rationale annotation}}
Incorporating additional argumentation for preference data is helpful to uncover more advanced patterns in human and LLMs behaviour, as well as useful in datasets for preference alignment \cite{just2024data, liu2025uncovering}, like a dataset curated in this work. Therefore, we collect rationale annotation for coding pairs with midpoint preference choice\footnote{Each pair is re-annotated for rationale by its original annotator.} (in this case, 2 and 3, as for \textit{Slightly prefer A/B}). Existing work either adopts free-text rationale \cite{gabriel2020artificial} or preset templates where the annotators fill in specific blanks \cite{zhi2025beyond}. We adopt both methodologies for completeness. To filter out unreliable annotators who may offload rationale writing to LLMs, we adopt a mix of metric checks, manual inspection, and comparison with popular vendor LLMs (see Appendix \ref{sec:rationale}). Overall, we obtain rationale annotation from 26 experts over 514 coding snippets.

\section{Evaluating LLMs' consistency over subjective preferences}\label{sec:eval_llm_consistency}

As a first experiment, we assess if LLMs can effectively understand subjective preferences when explicitly expressing them. Given a Python problem $z$ from our dataset, we ask the LLM to choose between two preference options $x$ or $y$ (e.g. extensive or minimal commenting), to solve it. We do this with $x,y$ represented as both text and code. Ideally, we expect the model to be consistent (e.g., "extensive commenting", and the snippet with more comments). This tells us to what extent LLMs recognize the code realization of a textual preference.

We note that our annotators are expected to have this ability, given their work experience. Regardless of this, they also demonstrate it in our qualification task with an acceptance rate of 100\% accuracy to proceed with the main annotation task. We also note that human coders, during their job, do not explicitly state a preference: they apply it implicitly. An expert AI assistant, on the other hand, should motivate its choices, and a developer may even state their own preference before, to condition them (e.g. “keep in mind I prefer more densely commented code”). 

We evaluate 13 LLMs (Appendix \ref{sec:annotationweb}) across four categories. As \textbf{baseline} models, we include Llama-3-8B-Instruct (\llamasmall) and Llama-3-70B-Instruct (\llamabig) \cite{grattafiori2024llama}, DeepSeek-R1-Distill-Qwen-14B (\deepseek), and \qwensmall and \qwenbig \cite{yang2025qwen3}. As \textbf{coding} models, we use \starcoder \cite{lozhkov2024starcoder}, DeepSeek-Coder-33B-Base (\deepseekcode) \cite{guo2024deepseek}, and Qwen3-Coder-30B-Instruct (\qwencoder) \cite{yang2025qwen3}. As \textbf{judge} models, we include \prometheus \cite{kim2024prometheus} and DeepSeek-GRM-16B (\deepseekgrm) \cite{liu2025inference}. Finally, we evaluate three \textbf{vendor} models: gpt-5-mini (\gpt) \cite{singh2025openai}, Claude-sonnet-4.5 (\claude) \cite{anthropic2025claudesonnet}, and gemini-3-flash-preview (\gemini) \cite{google2025gemini}. We test three setups:


\paragraph{\textbf{Baseline (BL)}:} LLM is asked to specify the preferences in Session A, where the model selects a textual option, and Session B, where the model selects a code snippet. Session B is new and holds no memory of Session A.
\paragraph{\textbf{One-chat (OC)}:} Both choices are made in the same session, ruling out forgetting the textual explanation that can be possible in the \textit{Baseline} setting. 
\paragraph{\textbf{Few-Shot (FS)}:} We introduce 2 input-output examples of paired textual and coding preferences to the prompt. 

The specific prompts can be found in Appendix \ref{sec:prompts}. Additionally, we re-run each setup by swapping the order of the two snippets to investigate positional bias. We calculate \textit{Consistency Score} ($\mathrm{CS}$) as the correspondence between preference text and code choices:

\begingroup\small
\[
\mathrm{CS} = P(Q_t, Q_c) \sim D\!\bigl(\mathrm{LLM}(Q_t) = \mathrm{LLM}(Q_c)\bigr)
\]
\endgroup

where $P(Q_t, Q_c)$ denotes the joint preference over a text query $Q_t$ and a code query $Q_c$, $\mathrm{LLM}(\cdot)$ is the response of the language model to a given query, and $D(\cdot)$ is an indicator of agreement between the two responses. When we swap the snippets, we calculate \textit{Swap Adjustment Score} ($\mathrm{SAS}$), measuring LLMs' consistency when the order of preferences is shifted. This metric denotes the percentage of cases where the model chooses the same option both in the ordered and the shifted dataset. Thus, such cases, when the choice for the first setup is \textit{A}, and the choice for the shifted setup is \textit{B}, which represent the same unit:

\begingroup\small
\[
\mathrm{SAS}
=
\Pr\!\big(
\mathrm{LLM}(Q_{\text{ordered}})
\neq
\mathrm{LLM}(Q_{\text{shifted}})
\big)
\]
\endgroup

where $Q_{\text{ordered}}$ and $Q_{\text{shifted}}$ denote the query with the original and swapped snippet ordering, respectively, and $\Pr(\cdot)$ is the probability over the evaluated cases.

\begin{table}[t]
\centering
\small
\resizebox{\columnwidth}{!}{%
\begin{tabular}{l*{3}{cc}}
\hline
\textbf{Open LLMs}
  & \multicolumn{2}{c}{BL}
  & \multicolumn{2}{c}{OC}
  & \multicolumn{2}{c}{FS} \\
\cline{2-7}
  & $\mathrm{CS}{\uparrow}$ & $\mathrm{SAS}{\uparrow}$
  & $\mathrm{CS}{\uparrow}$ & $\mathrm{SAS}{\uparrow}$
  & $\mathrm{CS}{\uparrow}$ & $\mathrm{SAS}{\uparrow}$ \\
\hline
\llamasmall     & 0.35 & 0.14 & 0.36 & 0.15 & 0.67 & 0.48 \\
\llamabig    & 0.68 & 0.55 & 0.67 & \cellcolor{green!15}0.63 & 0.72 & 0.63 \\
\qwensmall     & 0.90 & \cellcolor{red!15}0.00 & \cellcolor{green!15}1.00 & \cellcolor{red!15}0.00 & 0.96 & \cellcolor{red!15}0.00 \\
\qwenbig    & 0.80 & 0.20 & \cellcolor{green!15}1.00 & \cellcolor{red!15}0.00 & 0.81 & \cellcolor{red!15}0.00 \\
\deepseek    & \cellcolor{green!15}0.92 & \cellcolor{red!15}0.00 & \cellcolor{green!15}1.00 & \cellcolor{red!15}0.00 & \cellcolor{green!15}0.99 & \cellcolor{red!15}0.00 \\
\deepseekcode & 0.53 & 0.02 & 0.73 & 0.04 & 0.73 & 0.17 \\
\starcoder   & 0.26 & 0.07 & 0.52 & 0.24 & 0.51 & 0.03 \\
\qwencoder  & 0.50 & 0.40 & 0.57 & 0.44 & 0.63 & 0.53 \\
\prometheus   & \cellcolor{red!15}0.11 & \cellcolor{red!15}0.00 & 0.52 & 0.08 & \cellcolor{green!15}0.99 & \cellcolor{red!15}0.00 \\
\deepseekgrm   & 0.49 & 0.13 & 0.80 & 0.17 & 0.45 & 0.04 \\
\hline
\end{tabular}}

\smallskip

\resizebox{\columnwidth}{!}{%
\begin{tabular}{l*{3}{cc}}
\hline
\textbf{Vendor LLMs}
  & \multicolumn{2}{c}{BL}
  & \multicolumn{2}{c}{OC}
  & \multicolumn{2}{c}{FS} \\
\cline{2-7}
  & $\mathrm{CS}{\uparrow}$ & $\mathrm{SAS}{\uparrow}$
  & $\mathrm{CS}{\uparrow}$ & $\mathrm{SAS}{\uparrow}$
  & $\mathrm{CS}{\uparrow}$ & $\mathrm{SAS}{\uparrow}$ \\
\hline
\gpt    & 0.25 & \cellcolor{green!15}0.63 & \cellcolor{red!15}0.21 & \cellcolor{green!15}0.61 & \cellcolor{red!15}0.29 & \cellcolor{green!15}0.65 \\
\claude & 0.27 & 0.35 & \cellcolor{red!15}0.15 & 0.57 & \cellcolor{red!15}0.18 & 0.25 \\
\gemini & 0.35 & 0.55 & 0.33 & 0.60 & 0.30 & \cellcolor{green!15}0.67 \\
\hline
\end{tabular}}
\caption{LLMs' $CS$ ($\uparrow$) and $SAS$ ($\uparrow$) across different prompting approaches. We mark {\colorbox{green!15}{\textbf{best}}} and {\colorbox{red!15}{\textbf{worst}}} per column.}
\label{tab:consistency}
\end{table}

\begin{figure*}[t]
\centering
\includegraphics[width=0.9\linewidth]{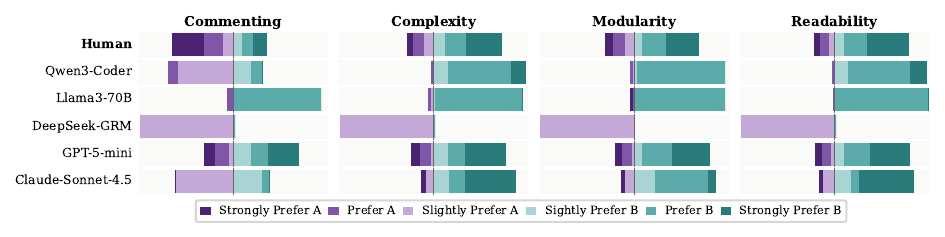}
  \caption {Distribution of preference choices among LLMs and humans.}
  \label{fig:behavior_plots}
\end{figure*}

\begin{table*}[t]
\centering
\small
\renewcommand{\arraystretch}{1.2}
\begin{tabular}{lcccccccccc}
\toprule
& \multicolumn{2}{c}{\textbf{Commenting}}
& \multicolumn{2}{c}{\textbf{Complexity}}
& \multicolumn{2}{c}{\textbf{Modularity}}
& \multicolumn{2}{c}{\textbf{Readability}}
& \multicolumn{2}{c}{\textbf{Overall}} \\
\cmidrule(lr){2-3}\cmidrule(lr){4-5}\cmidrule(lr){6-7}\cmidrule(lr){8-9}\cmidrule(lr){10-11}
\textbf{Human VS}
& $D$ & $JSD$ & $D$ & $JSD$ & $D$ & $JSD$ & $D$ & $JSD$ & $D$ & $JSD$ \\
\midrule
Human        & 1.93& --   & 1.67    & --   & 1.81    & --   & 1.49    & --   & 1.74    & --   \\
\midrule
\llamabig    & \cellcolor{red!15}2.29    & 0.63 & \cellcolor{green!15}1.30    & 0.55 & 1.36    & 0.56 & \cellcolor{green!15}1.18    & 0.60 & \cellcolor{green!15}1.54    & 0.57 \\
\qwencoder   & 1.85    & 0.51 & 1.09 & 0.38 & \cellcolor{green!15}1.05    & 0.54 & 1.02    & 0.36 & 1.25    & 0.39 \\
\deepseekgrm & 1.64    & \cellcolor{red!15}0.70 & \cellcolor{red!15}1.92 & \cellcolor{red!15}0.71 & \cellcolor{red!15}1.95 & \cellcolor{red!15}0.73 & \cellcolor{red!15}2.12 & \cellcolor{red!15}0.74 & \cellcolor{red!15}1.89 & \cellcolor{red!15}0.72 \\
\gpt         & \cellcolor{red!15}2.31    & \cellcolor{green!15}0.26 & 1.71 & \cellcolor{green!15}0.12 & 1.79 & \cellcolor{green!15}0.12 & 1.56 & \cellcolor{green!15}0.04 & 1.85 & \cellcolor{green!15}0.12 \\
\claude      & 1.71    & 0.58 & 1.45    & 0.21 & 1.46    & 0.37 & 1.29    & 0.26 & 1.49    & 0.28 \\
\bottomrule
\end{tabular}
\caption{Per-metric disagreement ($D$ $\downarrow$) and Jensen-Shannon divergence ($JSD$ $\downarrow$) between annotator pairs. We mark {\colorbox{green!15}{\textbf{best}}} and {\colorbox{red!15}{\textbf{worst}}} values for both $D$ and $JSD$ across metrics.}
\label{tab:alignment_combined}
\end{table*}

From results (\Cref{tab:consistency}), models consistently pick contradicting textual and code preferences. This is evident from the generally low $CS$ values. When a very high $CS$ appears, this is due to positional bias. For example, \qwensmall, \qwenbig, and \deepseek reach $CS > 80$. However, once we swap the order of the snippets, the models do not change their choice, and $SAS$ drops dramatically. Interestingly, $CS$ and $SAS$ show opposite trends in \gpt and \gemini, showing more positional consistency in their choices. We also note that OC and FS prompting do not improve the situation: some models benefit from it in terms of $CS$, but with little-to-no change in $SAS$.

\section{Exploring Preference Behavior among LLMs and Humans}\label{sec:preference_behavior_llms_humans}

Next, we compare the distribution of preference values between humans and LLMs. To manage the computational and inference cost of our experiments, we select 3 open-weight and 2 vendor LLMs based on combined $CS$ and $SAS$ (\Cref{tab:consistency}), which is calculated as a summary of these two metrics in the Baseline setup. This is motivated by the intuition that models should have high $CS$ and high $SAS$ to avoid positional bias. We choose \llamabig (baseline model); \qwencoder (coding model); \deepseekgrm (judge model); \claude and \gpt (vendor models). We prompt these models by replicating our human annotation setup (\Cref{sec:human_preference_annotation}), meaning that LLMs are now expressing a preference on a 6-scale, given the two snippets and a description of the axis value.

\paragraph{\textbf{Disagreement}} We first compute the distribution of human Likert rankings from \Cref{sec:human_preference_annotation}. For each coding pair $i$, we define a per-item disagreement score $a_i$: for human annotators, it is the average pairwise absolute difference across all $\binom{n_i}{2}$ annotator pairs $(j,k)$, where $h_{ij}$ and $h_{ik}$ are their respective Likert ratings; for an LLM, it is the absolute deviation of its response $\ell_i$ from the human mean $m_i$. The overall Disagreement $D$ is then the mean of $a_i$ over all $N$ coding pairs:

\begingroup\small
\begin{align*}
a_i &=
\begin{cases}
\dfrac{1}{\binom{n_i}{2}}
\displaystyle\sum_{1 \leq j < k \leq n_i}
\left| h_{ij} - h_{ik} \right|
& \text{for human annotation} \\[10pt]
\left| \ell_i - m_i \right|
& \text{for LLM annotation}
\end{cases}
\end{align*}
\begin{equation*}
D = \frac{1}{N} \sum_{i=1}^{N} a_i
\end{equation*}
\endgroup

Via this metric, we compute  $D$ among humans, and between humans and LLMs. We also compute Jensen-Shannon Divergence~\citep{lin1991divergence} between humans and LLMs. We choose this measure over the others (such as KL divergence~\citep{kullback1951information}) as it is symmetric and bounded, offering better interpretation of its values. This is in line with other works in the field~\citep{go2023aligning,wang2024beyond}. We intentionally avoid metrics for measuring inter-annotator agreement, as this work focuses on \textit{subjective} and targeting a gold standard for per-item LLM comparison would lead to a paradox. 

Results from the distribution (\Cref{fig:behavior_plots}) and metrics (\Cref{tab:alignment_combined}) reveal how critical it is to assess alignment via multiple metrics. \gpt achieves the lowest $JSD$ scores, with a distribution of ratings closely mirroring the experts (except for commenting). However, it shows one of the highest $D$: despite producing realistic aggregate statistics, \gpt still frequently deviates from human judgement on individual examples. This indicates that a model can learn the "shape" of human opinions without learning what drives human decisions on specific cases. \llamabig shows the opposite failure mode: systematic distribution biases ($JSD$), but low $D$. This shows how a consistent directional bias can accidentally align with human medians. Generally, all open LLMs show extremely skewed distribution; \deepseekgrm consistently shows the worst results for each metric, which could be attributed to it being the smallest model. Additional statistical analysis of these results is in Appendix \ref{sec:bootstrapping}. 

\paragraph{\textbf{Qualitative analysis}} Subjectivity can be trivial for judgments on Commenting and Modularity. Computational complexity, however, includes more intricate factors like worst-case complexity, expected input distribution, amortized complexity, or computational limitations. To understand the sources of disagreement along this axis, we perform a qualitative analysis of the experts' judgments compared to \gpt, as it shows the strongest $JSD$ (\Cref{tab:alignment_combined}). We inspect 20 randomly selected code pairs with at least one human annotator disagreeing with \gpt. First, we observe {\em reliance on external assumptions}: \gpt\ (alongside some experts) occasionally presumes additional contexts beyond the problem statement; for example, it favours code with better average case complexity, while experts may consider worst-case complexity (Appendix \ref{app:examples_qualitative_analysis}, ID 1). 
Such hidden contexts of preferences can misdirect preference-learning \cite{siththaranjan2024distributional}, and are of utmost importance in case of expert preferences. Similarly, \gpt sometimes relies on auxiliary metrics beyond complexity, like commenting, readability, pure-pythonic vs mixed-implementations, and usage of certain libraries such as Numpy (\ref{app:examples_qualitative_analysis}, ID 2-4). 
This sometimes leads to choosing an inefficient implementation due to surface-level simplicity. Finally, we observe consistent {\em brittle reasoning} in \gpt: its rationale (\ref{app:examples_qualitative_analysis}, ID 5) 
relies on superficial characteristics of the code, or even blatantly ignores known algorithmic complexity bounds, echoing prior findings \cite{csahinucc2026reward}.

\section{Conclusions}
In this work, we inspected the behaviour of LLMs over subjective coding preferences, which are critical in expert coding, and often overlooked by traditional benchmarking. We introduced a novel dataset of coding snippets, that are expert-annotated with subjective preferences. We then used our dataset to identify patterns of behavior and consistency in LLMs. We find that LLMs cannot consistently identify the realization of subjective preferences, and struggle in identifying the same preference value when it's presented as text or code. We also find severe positional bias, showing that even seemingly robust models just learned fixed selection patterns. Vendor models are less affected by positional bias, but still fall far from desirable consistency. Compared to experts, most tested models show extremely polarized distribution, opting almost exclusively for the extreme of the proposed scale. \gpt closely resembles the human distribution, but still suffers from consistent per-item divergence. Our qualitative study further reveals problematic patterns in how the model reasons.

We release our dataset to foster research in preference learning and expert coder models. Our dataset can be used to integrate coding preferences into human reinforcement strategies for further development of coding assistants that follow users' personal choices. Additionally, acquired annotations and rationales can serve as a source for extended linguistic and statistical analysis for deeper insights into how LLMs understand code.

\section{Limitations}

\paragraph{\textbf{Preference Matching and Preference Aggregation}} 
We focus on four coding preference axes. We do understand, however, that subjective preferences may widen into many other axes we did not consider. Some examples of those from the related literature and results of our survey include code \textit{maintainability} and \textit{security}. We also acknowledge that preference axes could aggregate in complex ways, entangling and co-influencing each other. Our current framework represents the single-principal alignment for an AI system to match one unique preference. We reserve the analysis of complex aggregation of multiple axes within single coding snippets for future work.

\paragraph{\textbf{Programming Languages}} 
We limit our study to Python, excluding other programming languages. This choice was motivated by the extensive set of  Python resources available online; however, other programming languages should be considered in future developments of our research line. 


\paragraph{\textbf{Model choices}} The choice of our 13 LLMs is certainly limited, especially when it comes to the rapid development of new models, scales, and training approaches. In this work, we restricted ourselves to those baseline, coding, judge, and vendor models that showed the strongest performance in popular coding benchmarks. This is a conscious choice that lets us analyze the models that are more likely to be used by software engineers, when it comes to both open and vendor LLMs. This was also a necessity to keep a manageable computational requirement and experimental runtime: testing each model on $\approx3000$ paired snippets, in three prompting setups, including the swapped case, imposed us to take this choice. The same logic made us narrow our selection in our second experiment, where we picked the 5 most consistent models. Additionally, this was motivated by the necessity of keeping the inference cost of vendor model within our budget (accounting also for the budge spent in human annotation).

\paragraph{\textbf{Assumption of expert competence}} We consider our coding experts as the gold standard when it comes to real-world subjective preferences. However, we do not assume that our experts are always correct. We acknowledge that our qualification task, annotation setup, and sanity checks may be limited and not fully sufficient against all the challenges that come into the implicitly complex domain of subjective data annotation. We assume the trustworthiness of our expert based on our protocol, but also acknowledge that the convoluted nature of expert preference may need deeper analysis.

\paragraph{\textbf{Application of the chosen binary scenario in coding assistants}} This work investigates the preliminary task and fundamental premise of the successful implementation of modern coding assistants. More specifically, contemporary agentic practices implicitly or explicitly rely on the agent's ability to understand what is preferred and or dispreferred. In the proposed setup, we directly evaluate if the LLM can understand the human’s input. However, we acknowledge the possible limitation of this proxy and do not state that it is a logically sufficient guarantee of the downstream competence of coding agentic systems. 

\bibliography{custom}

\appendix

\section{Additional information on coding metrics}\label{section:readability_metric}

\begin{figure}[t]
  \centering
  \includegraphics[width=\columnwidth]{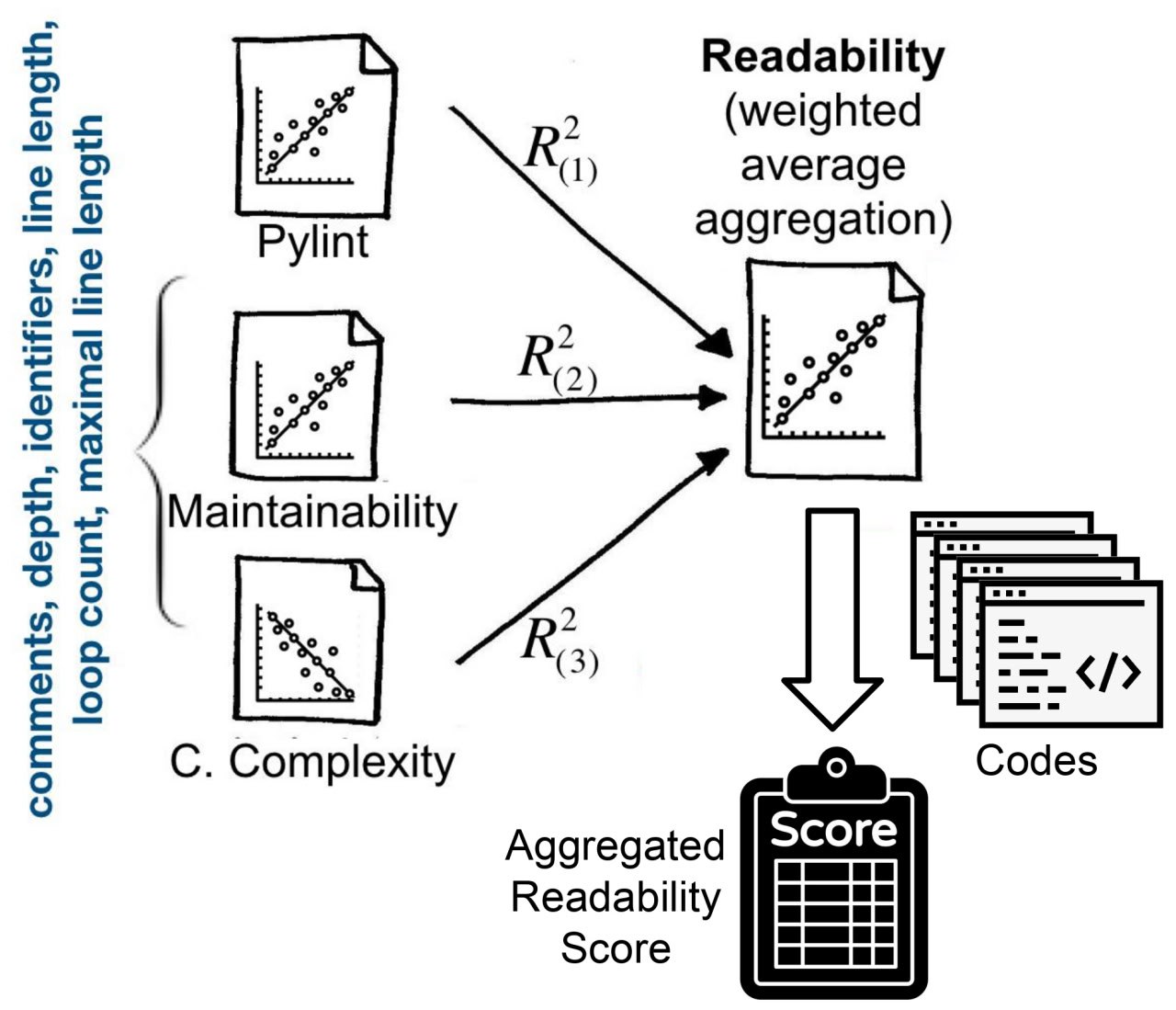}
  \caption{Readability Model Pipeline.}
  \label{fig:readability}
\end{figure}

Although there is no single objective criteria for readability, discussions on this topic agree on the following characteristics that are usually associated with how easy code is to understand:
(1) Following coding standards, linters, pep8, etc.;
(2) Cyclomatic complexity, which is a way of evaluating the code based on a built graph and number of nodes, edges and connected components;
(3) Code maintainability, including above-mentioned cyclomatic complexity as well as relations between operators and operands;
(4) Subjective human perception of how code looks which is based on local factors. These are the surface features that aim to capture human notions of readability and include such traits as number of comments, average line length, and loop count.
We propose creating a combined readability model (Figure \ref{fig:readability}) that, on one hand, contains overt code features, and, on the other hand, represents the established concepts of readability. For this purpose, a multiple linear regression is used with code local factors as independent variables and scores acquired by three methods that to some extent reflect readability as dependent variables on this data \cite{maheshwari_code_snippets_readability}. We then obtain weights for each of the local features, which allow us to reproduce the readability metric on any code snippet of any programming language. To filter code pairs from the dataset for the readability samples, we selected instances that were either more readable and less complex or less readable and more complex, using the above-described metric.

\begin{figure}[t]
  \centering
  \includegraphics[width=\columnwidth]{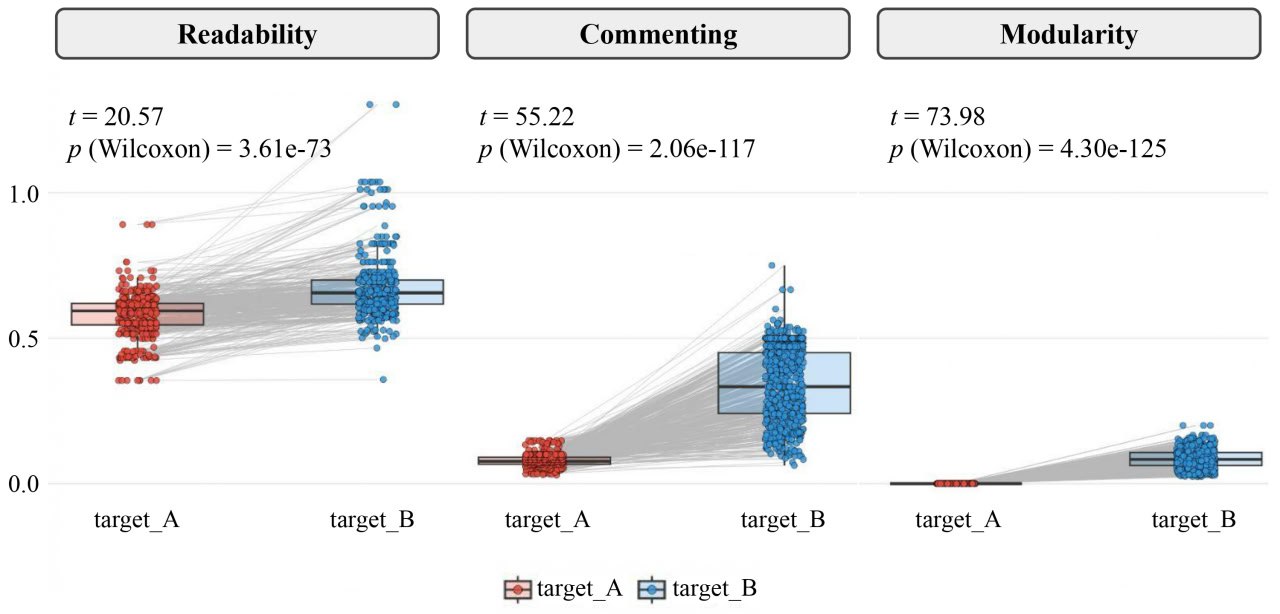}
  \caption{Distribution of code metrics differentiating Code\_A and Code\_B.}
  \label{fig:dataset_variety}
\end{figure}

\section{Ethics Statement}\label{app:ethics_statement}

In this section, we briefly discuss the ethical aspects of our experiments.

\paragraph{\textbf{Ethical Review}} Before starting our experiments, we consulted our Institution for ethical pre-assessment. Since we do not collect any demographics and personal information about the participants, and crowdworkers do not disclose any data that could identify them, our experiment was confirmed to be ethically compliant without further procedures.

\paragraph{\textbf{Data Consent}} For both recruitment processes done in this study, the survey and the main annotation task, we provided instructions explaining how the collected data would be used, the goal of the study, and ways to contact the study executors directly. Participants who wanted to take part in our experiments were required to give us explicit agreement before proceeding. 

\paragraph{\textbf{Workload and pay}} We estimated a maximum time for completing the expert survey of 30 min, and set an hourly rate per participant of 22€. 

For the numeric preference annotation, we first conducted a pilot study, where we determined the average completion time as one hour for \textasciitilde90 coding pairs. For the qualification task, which consisted of annotating 8 coding pairs, the required time per participant was 15 min and the hourly rate was 28€. For the full-scale annotation task, the hourly rate per annotator was set to 14€.

For the rationale preference annotation, participants are paid approximately 15€/hour, and the estimated (from the pilot study) workload is 6 coding pairs per hour.

\paragraph{\textbf{Additional information on the recruited experts}} 
Other general information about participants of the \textit{survey} includes:
\begin{enumerate}
    \item 44\% have 4-6 years of working experience; 28\% have 7-9 years; 20\% have 1-3 years; and 8\% have 10+ years.
    \item 48\% spend approximately 3-5 hours a day on working with code; 44\% spend 5-8 hours a day; 8\% spend 1-3 hours.
\end{enumerate}

\section{Survey Coding Styles of Software Engineers}\label{app:survey}

\newtcolorbox{qbox}{
  colback=gray!3, colframe=gray!40,
  boxrule=0.5pt, arc=3pt,
  left=8pt, right=8pt, top=6pt, bottom=6pt
}

\newcommand{\Q}[2]{\begin{qbox}\textbf{Q#1.} #2\end{qbox}}
\newcommand{\Resp}{\noindent\textit{Response options:}\ }

\newtcolorbox{sectionbox}{
  colback=gray!6,
  colframe=gray!40,
  boxrule=0.6pt,
  arc=3pt,
  left=10pt, right=10pt, top=7pt, bottom=7pt
}

\newcommand{\SurveySection}[1]{%
  \begin{sectionbox}
    \normalsize\bfseries #1
  \end{sectionbox}
}

\SurveySection{Part 1: General information about experience and current work}

\Q{1}{How many years of working experience do you have as a developer?}
\Resp
\begin{enumerate}[label=(\alph*), leftmargin=*, nosep]
  \item 1-3 years
  \item 4-6 years
  \item 7-9 years
  \item 10+ years
\end{enumerate}

\Q{2}{What is your main specialisation?}
\Resp
\begin{enumerate}[label=$\square$, leftmargin=*, nosep]
  \item Back-end
  \item Front-end
  \item Full Stack
  \item QA/testing
  \item ML engineering
  \item Other: \underline{\hspace{3cm}}
\end{enumerate}

\Q{3}{What are the main applications of your code?}
\Resp
\begin{enumerate}[label=(\alph*), leftmargin=*, nosep]
  \item Company projects
  \item Academic projects
\end{enumerate}

\Q{4}{How many hours a day do you spend on working with code on average?}
\Resp
\begin{enumerate}[label=(\alph*), leftmargin=*, nosep]
  \item 1-3 hours
  \item 3-5 hours
  \item 5-8 hours
\end{enumerate}

\SurveySection{Part 2: Coding Comments}

\textit{Instructions:}

You will be asked to rate several coding factors based on how relevant they are in your daily work and your personal experience. For your convenience and clearness, each coding factor is provided with a short definition.

"Coding comments" refer to the commented or docstring lines in code that usually explain what the code does

Example case: some code might contain a high number of comments explaining each step, while another way to implement the same code is with a moderate use of comments in the exceptional cases.

\Q{1}{To what extent do you agree with the following statement regarding the use of coding comments?

1. Using different degree of comments in code (i.e. low or high) depends on the habits of a particular developer

2. In my daily coding activity I personally choose different degree of comments (i.e. low or high) depending on the specific project/task

3. Using different degree of comments in code depends on the requirements and/or computational limitations of the specific project

4. Using different degree of comments in code (i.e. low or high) doesn't depend on the habits of a particular developer}
\Resp 1 = Strongly disagree,\ 2 = Disagree,\ 3 = Somewhat disagree,\ 4 = Neutral,\ 5 = Somewhat agree,\ 6 = Agree,\ 7 = Strongly agree.

\Q{2}{Please mark here how you usually balance degree of comments in your code}
\Resp
\begin{enumerate}[label=(\alph*), leftmargin=*, nosep]
  \item I always use minimal commenting
  \item I mostly use minimal commenting
  \item I equally use minimal and extensive commenting
  \item I mostly use extensive commenting
  \item I always use extensive commenting
\end{enumerate}

\Q{3}{Feel free to share any other thoughts about this coding characteristic (coding comments)}
\Resp Open-ended text.
\vspace{1.2em}

\SurveySection{Part 3: Coding Modularity}

\textit{Instructions:}

You will be asked to rate several coding factors based on how relevant they are in your daily work and your personal experience. For your convenience and clearness, each coding factor is provided with a short definition.

"Modularity" refers to the higher or lower usage of smaller self-contained modules (functions and classes)

Example case: some code might contain higher number of smaller self-contained modules, while another way to implement the same scripts is to write monolithic code

\Q{1}{To what extent do you agree with the following statement regarding the use of coding modularity?

1. Modularity in code depends on the requirements and/or computational limitations of the specific project

2. Modularity in code doesn't depend on the habits of a particular developer

3. In my daily coding activity I personally choose different degree of modularity depending on the specific project/task

4. Modularity in code depends on the habits of a particular developer}
\Resp 1 = Strongly disagree,\ 2 = Disagree,\ 3 = Somewhat disagree,\ 4 = Neutral,\ 5 = Somewhat agree,\ 6 = Agree,\ 7 = Strongly agree.

\Q{2}{Please mark here how you usually balance modularity in your code}
\Resp
\begin{enumerate}[label=(\alph*), leftmargin=*, nosep]
  \item I always write monolithic code
  \item I mostly write monolithic code
  \item I equally write monolithic code and code that contains higher number of smaller modules
  \item I mostly write code that contains higher number of smaller modules
  \item I always write code that contains higher number of smaller modules
\end{enumerate}

\Q{3}{Feel free to share any other thoughts about this coding characteristic (modularity)}
\Resp Open-ended text.
\vspace{1.2em}

\SurveySection{Part 4: Code Complexity}

\textit{Instructions:}

You will be asked to rate several coding factors based on how relevant they are in your daily work and your personal experience. For your convenience and clarity, each coding factor is provided with a short definition.

"Complexity" refers to the balance between computational resources and the amount of time taken by the code to run.

Example case: some code might be very quick to execute but requires a lot of memory, while another way to implement the same code is to save memory and sacrifice the running time

\Q{1}{To what extent do you agree with the following statement regarding the use of coding complexity?

1. Complexity in code depends on the habits of a particular developer

2. Complexity in code depends on the requirements and/or computational limitations of the specific project

3. In my daily coding activity I personally make decisions regarding code complexity depending on the specific project/task

4. In my daily coding activity my decisions regarding code complexity don't depend on the specific project/task}
\Resp 1 = Strongly disagree,\ 2 = Disagree,\ 3 = Somewhat disagree,\ 4 = Neutral,\ 5 = Somewhat agree,\ 6 = Agree,\ 7 = Strongly agree.

\Q{2}{Please mark here how you usually balance complexity in your code}
\Resp
\begin{enumerate}[label=(\alph*), leftmargin=*, nosep]
  \item I always minimize computational resources and sacrifice speed of code execution
  \item I mostly minimize computational resources and sacrifice speed of code execution
  \item I equally balance computational resources and speed of code execution
  \item I mostly minimize speed of code execution and sacrifice computational resources
  \item I always minimize speed of code execution and sacrifice computational resources
\end{enumerate}

\Q{3}{Feel free to share any other thoughts about this coding characteristic (complexity)}
\Resp Open-ended text.
\vspace{1.2em}

\SurveySection{Part 5: Code Readability}

\textit{Instructions:}

You will be asked to rate several coding factors based on how relevant they are in your daily work and your personal experience. For your convenience and clearness, each coding factor is provided with a short definition.

"Readability" refers to how easy code is to understand, "readability-complexity trade-off" implies the balance between how readable a code is and the computational costs to run it

Example case: some very optimised code might be very hard to read without deep understanding of the underlying framework, while another way to implement the same code is to sacrifice computational costs for better code readability

\Q{1}{To what extent do you agree with the following statement regarding the use of coding readability-complexity trade-off?

1. Choices regarding coding readability-complexity trade-off depend on the requirements and/or computational limitations of the specific project

2. In my daily coding activity I personally make decisions regarding readability-complexity trade-off depending on the specific project/task

3. Choices regarding coding readability-complexity trade-off don't depend on the requirements and/or computational limitations of the specific project

4. Choices regarding coding readability-complexity trade-off depends on the habits of a particular developer}
\Resp 1 = Strongly disagree,\ 2 = Disagree,\ 3 = Somewhat disagree,\ 4 = Neutral,\ 5 = Somewhat agree,\ 6 = Agree,\ 7 = Strongly agree.

\Q{2}{Please mark here how you usually balance readability-complexity trade-off in your code}
\Resp
\begin{enumerate}[label=(\alph*), leftmargin=*, nosep]
  \item I always sacrifice computational resources for a more readable code
  \item I mostly sacrifice computational resources for a more readable code
  \item I equally balance computational resources and how readable code is
  \item I mostly sacrifice how readable code is and minimize computational resources
  \item I always sacrifice how readable code is and minimize computational resources
\end{enumerate}

\Q{3}{Feel free to share any other thoughts about this coding characteristic (readability-efficiency trade-off)}
\Resp Open-ended text.
\vspace{1.2em}

\SurveySection{Part 6: Conclusion}

\Q{1}{What are the other coding factors that you’re considering important at your daily work?}
\Resp Open-ended text.
\vspace{1.2em}

\Q{2}{What are the other trade-offs in your daily coding decisions that you face? (i.e. produce memory-efficient code but sacrifice time of execution)}
\Resp Open-ended text.
\vspace{1.2em}

\Q{3}{Please share any other thoughts or feedback about different coding styles and choices that developers come across to}
\Resp Open-ended text.
\vspace{1.2em}

\section{Prompts}\label{sec:prompts}
We report the prompts we used in our experiments in \Cref{fig:mainprompt_template,figure:prompts_codegemma,figure:few_shot_prompts,figure:prompt_human_setup}. 

The selected prompts follow previous research on LLM preference among words and actions in different domains \cite{li2024dissecting}. The selected prompts do not contain any distinctive stylistic features that would drastically influence LLMs' outputs. For example, none of the prompts has an articulated phrasing that is unique to the axes implementation. Nevertheless, the results remain consistent across model families, model sizes, and prompting configurations. This consistency strengthens the reliability of the findings, since it is unlikely that a specific prompt formulation would produce highly similar outcomes across all these conditions. Moreover, the majority of each prompt consists of fixed snippet pairs. Given the autoregressive nature of large language models, this further suggests that the comparatively small variable portion of the prompt is unlikely to substantially alter the model’s behaviour.

\begin{figure}[!htb]
\begin{minipage}{0.49\textwidth}
\promptbox{promptbg}{promptframe}{promptfont}{Prompt template to investigate LLMs' coding preferences. $x$ and $y$ are realized either as textual descriptions or code snippets solving a Python problem $z$.}{fig:mainprompt_template}{1}{\tiny\texttt{The following two options represent solutions of the same Python problem $z$.Which of these two options would you prefer?\\~\\
Option A: $x$\\~\\
Option B: $y$\\~\\
Please respond with only A or B.}}
\end{minipage}
\end{figure}

\begin{figure}[!htb]
\begin{minipage}{0.49\textwidth}
\promptbox{promptbg}{promptframe}{promptfont}{CodeGemma-7B Code Generation Prompt.}{figure:prompts_codegemma}{1}{\tiny\texttt{Here is a code snippet \textbf{\$CODE} that solves the following Python problem: \textbf{\$PROBLEM}. This code scores very low on \textbf{\$PREFERENCE}.\\~\\
Write the opposite of this code solution for the stated problem so that your version will score high on \textbf{\$PREFERENCE}\\~\\
Here is the explanation of \textbf{\$PREFERENCE}: \textbf{\$EXPLANATION}}}
\end{minipage}
\end{figure}

\begin{figure}[!htb]
\begin{minipage}{0.49\textwidth}
\vspace{5mm}
\promptbox{promptbg}{promptframe}{promptfont}{Few-Shot Example Prompt.}{figure:few_shot_prompts}{1}{\tiny\texttt{The following two options represent solutions of the same Python problem $z$.Which of these two options would you prefer?\\~\\
Option A: $x$\\~\\
Option B: $y$\\~\\
Please respond with only A or B.\\~\\
Here are some examples:\\~\\
1) Option A \textbf{\$CODE}: $x_{code}$;
Option B \textbf{\$CODE}: $y_{code}$; Response:  $x_{code}$.
Option A \textbf{\$TEXT}: $x_{text}$;
Option B \textbf{\$TEXT}: $y_{text}$; Response:  $x_{text}$.\\~\\
2) Option A \textbf{\$CODE}: $x_{code}$;
Option B \textbf{\$CODE}: $y_{code}$; Response:  $y_{code}$.
Option A \textbf{\$TEXT}: $x_{text}$;
Option B \textbf{\$TEXT}: $y_{text}$; Response:  $y_{text}$.}}
\end{minipage}
\end{figure}

\begin{figure}[!htb]
\begin{minipage}{0.49\textwidth}
\vspace{5mm}
\promptbox{promptbg}{promptframe}{promptfont}{Prompt for Human-like Preference Annotation.}{figure:prompt_human_setup}{1}{\tiny\texttt{\textbf{\$PROBLEM\_TITLE}\\~\\
\textbf{\$PROBLEM\_EXPLANATION}\\~\\
Which of these two codes would you most likely choose in a daily work as software engineer?\\~\\
Preference (0 = strongly prefer A, 5 = strongly prefer B)\\~\\
\textbf{\$PREFERENCE\_EXPLANATION}\\~\\
Option A: \textbf{\$TEXT\_A; \$CODE\_A}\\~\\
Option B: \textbf{\$TEXT\_B; \$CODE\_B}
}}
\end{minipage}
\end{figure}






\section{Details on Dataset Curation and CodeGemma-7B Applications}\label{section:codegemma7b}

In this section, we highlight that CodeGemma-7B did not generate the alternative code, but rather a modification that was manually checked to eliminate the potential differences that can be detected as “LLM-generated vs. human-written”. Thus, the LLM is used as a verification tool, whereas we are not building a direct connection, stating that judgment is harder than verification. We also validate the generated snippets through controlled runs and compare their output against the starting ones. We obtain only $\approx 14\%$ of snippets with mismatched stderror outputs. Upon manual review, we discover that this is not due to logical or coding inconsistencies, but rather from randomly generated test variables or minor formatting differences. Therefore, we keep all the generated snippets. Considering the amount of manual work for CodeGemma-7B is also not considered a golden standard model for code generation. To show that CodeGemma-7B still fails at recognizing different preferential modalities, we extend our analysis in Table \ref{tab:consistency}. The results show that the model replicated the inconsistent behavior observed by other models (Table \ref{tab:consistency_2}). 

\begin{table}[t]
\centering
\small
\resizebox{\columnwidth}{!}{%
\begin{tabular}{l*{3}{cc}}
\hline
\textbf{Open LLMs}
  & \multicolumn{2}{c}{BL}
  & \multicolumn{2}{c}{OC}
  & \multicolumn{2}{c}{FS} \\
\cline{2-7}
  & $\mathrm{CS}{\uparrow}$ & $\mathrm{SAS}{\uparrow}$
  & $\mathrm{CS}{\uparrow}$ & $\mathrm{SAS}{\uparrow}$
  & $\mathrm{CS}{\uparrow}$ & $\mathrm{SAS}{\uparrow}$ \\
\hline
\llamasmall     & 0.35 & 0.14 & 0.36 & 0.15 & 0.67 & 0.48 \\
\llamabig    & 0.68 & 0.55 & 0.67 & \cellcolor{green!15}0.63 & 0.72 & 0.63 \\
\qwensmall     & 0.90 & \cellcolor{red!15}0.00 & \cellcolor{green!15}1.00 & \cellcolor{red!15}0.00 & 0.96 & \cellcolor{red!15}0.00 \\
\qwenbig    & 0.80 & 0.20 & \cellcolor{green!15}1.00 & \cellcolor{red!15}0.00 & 0.81 & \cellcolor{red!15}0.00 \\
\deepseek    & \cellcolor{green!15}0.92 & \cellcolor{red!15}0.00 & \cellcolor{green!15}1.00 & \cellcolor{red!15}0.00 & \cellcolor{green!15}0.99 & \cellcolor{red!15}0.00 \\
\deepseekcode & 0.53 & 0.02 & 0.73 & 0.04 & 0.73 & 0.17 \\
\starcoder   & 0.26 & 0.07 & 0.52 & 0.24 & 0.51 & 0.03 \\
\qwencoder  & 0.50 & 0.40 & 0.57 & 0.44 & 0.63 & 0.53 \\
\prometheus   & \cellcolor{red!15}0.11 & \cellcolor{red!15}0.00 & 0.52 & 0.08 & \cellcolor{green!15}0.99 & \cellcolor{red!15}0.00 \\
\deepseekgrm   & 0.49 & 0.13 & 0.80 & 0.17 & 0.45 & 0.04 \\
\textbf{CodeGemma-7B}   & 1.00 & 0.00 & 0.30 & 0.00 & 0.76 & 0.32 \\
\hline
\end{tabular}}

\smallskip

\resizebox{\columnwidth}{!}{%
\begin{tabular}{l*{3}{cc}}
\hline
\textbf{Vendor LLMs}
  & \multicolumn{2}{c}{BL}
  & \multicolumn{2}{c}{OC}
  & \multicolumn{2}{c}{FS} \\
\cline{2-7}
  & $\mathrm{CS}{\uparrow}$ & $\mathrm{SAS}{\uparrow}$
  & $\mathrm{CS}{\uparrow}$ & $\mathrm{SAS}{\uparrow}$
  & $\mathrm{CS}{\uparrow}$ & $\mathrm{SAS}{\uparrow}$ \\
\hline
\gpt    & 0.25 & \cellcolor{green!15}0.63 & \cellcolor{red!15}0.21 & \cellcolor{green!15}0.61 & \cellcolor{red!15}0.29 & \cellcolor{green!15}0.65 \\
\claude & 0.27 & 0.35 & \cellcolor{red!15}0.15 & 0.57 & \cellcolor{red!15}0.18 & 0.25 \\
\gemini & 0.35 & 0.55 & 0.33 & 0.60 & 0.30 & \cellcolor{green!15}0.67 \\
\hline
\end{tabular}}
\caption{LLMs' $CS$ ($\uparrow$) and $SAS$ ($\uparrow$) across different prompting approaches. We mark {\colorbox{green!15}{\textbf{best}}} and {\colorbox{red!15}{\textbf{worst}}} per column.}
\label{tab:consistency_2}
\end{table}

\section{Validating dataset variety}\label{section:validating_variety}

To inspect subjective preferences in clearly distinguishable scenarios, it is important that our dataset contains sufficiently different coding snippets for each pair. We verify this via a paired Wilcoxon signed-rank test based on the level of commenting (t=$55.22$; mean-diff=$0.260$), modularity (t=$73.98$; mean-diff=$0.086$), and readability (t=$20.57$; mean-diff=$0.098$). We obtain P-values $< 2.2e-16$ for each preference type, proving significant differences (Figure \ref{fig:dataset_variety}). For complexity, the paired t-test was not performed, as these samples are measured in big-O notation, so the benchmarking values are not numerical.

\section{Annotation Interface}\label{sec:annotationweb}

\begin{figure*}[t]
    \centering
  \includegraphics[width=0.85\linewidth]{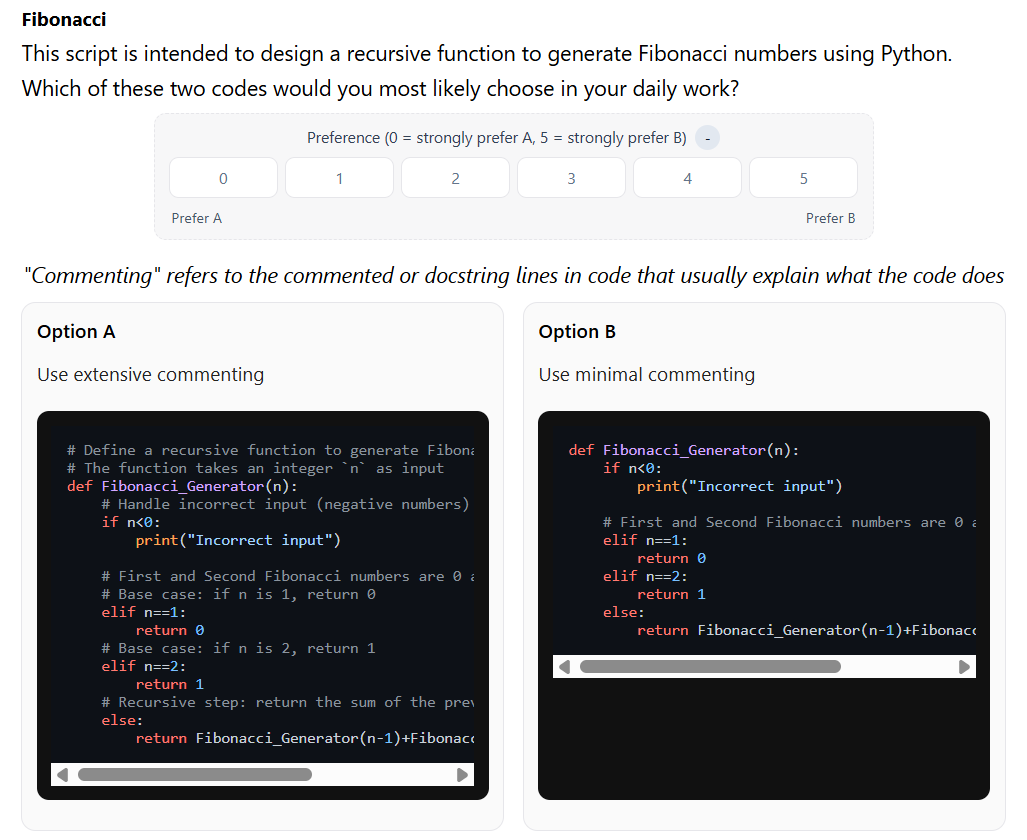}
  \caption {A full example of the expert preference annotation interface.}
  \label{fig:annotation}
\end{figure*}

This section demonstrates an example of UI interface for the main annotation task as it was displayed to participants (Figure \ref{fig:annotation}).

\section{Sanity checks}
\label{sec:rationale}

\paragraph{\textbf{Rationale annotation sanity check}} Before acquiring rationale from experts, we conduct a short probing rationale annotation study to understand the potential risks and limits for this kind of experimental setup. We contacted 4 experts who had already done the numeric annotation task, and asked them to ground their previous choices in the written form (2 coding pairs per each annotator). We compare the obtained answers with the responses from ChatGPT and Gemini by BertScore \cite{zhang2019bertscore}, as well as review them manually, to eliminate the use of LLMs for this round of annotation. This verification showed that both open-ended questions and a template can serve as a reliable source for collecting human rationales. In addition, we apply a combination of metrics on the sequences of numeric annotations for each participant to exclude lazy workers or participants whose previous annotations have been done mechanically. Those metrics include: entropy, switch rate (how often an annotator changes their label), switch symmetry (whether the annotator switches symmetrically and does random left-right clicking), percentage of using two midpoint labels together and separately, and frequency of the most repeated annotation sequence. We use the following thresholds for each metric: entropy $< 1.43$; switch rate $< 80\%$; switch symmetry $< 0.375$; usage of each midpoint label separately $< 23\%$; usage of midpoint labels together $< 41\%$.

\section{Experimental Setup}\label{app:exerimental_details}

The experimental setup for pairwise and 6-scale preference setting uses a fixed random seed across Python, NumPy, and PyTorch, and enables deterministic CUDA behavior to ensure reproducibility. For some models of larger size (e.g., \llamabig, \qwenbig, and \deepseekcode), inference is run with 8-bit quantization. We use greedy decoding (\texttt{do\_sample=False}) for deterministic token selection. The remaining settings, such as left padding and a maximum of new generated tokens, are standard implementation choices motivated by enforcing either binary-choice or N-token generation output. In some cases, the \texttt{max\_new\_tokens} hyperparameter was heuristically increased to avoid truncated outputs and ensure the model produced a complete decision.

\section{Statistics on Preference Behaviour among LLMs and Humans}\label{sec:bootstrapping}

In this section, we report the results of a bootstrap analysis with 1000 resamples of code pairs, including confidence intervals, standard errors, and cross-LLM as well as LLM-human comparisons for Table \ref{tab:alignment_combined}. For this analysis, we added three worst-performing models across three model families based on the summary of $CS$ and $SAS$ scores, which were \llamasmall, \starcoder, and \prometheus.

Analysis showed that across all three evaluation metrics - Disagreement (D), Jensen–Shannon (JS) divergence, and Spearman rank correlation - no model fully replicated human rating behaviour. Overall, the alignment metrics appear stable: standard errors are small ($\leq$ 0.019 for Disagreement, $\leq$ 0.018 for Spearman correlation), means match the original estimates to four decimal places, and biases are negligible ($\leq$ 0.003), indicating that results are not driven by a few items.

For Disagreement, or Mean Absolute Distance metric, Benjamini–Hochberg (BH)-adjusted paired bootstrap tests (two-sided) indicated that the vast majority of model pairings were significantly differentiated at the p < 0.01 level across the overwhelming majority of dimensions, with the small number of pairwise comparisons failing to reach significance (Claude versus GPT-5-mini in readability (p = 0.850) and Qwen3-Coder-30B versus Llama-3-70B in modularity (p = 0.376)). Moreover, pairwise differences between best- and worst-performing models were large in magnitude. For example, the overall difference between Qwen3-Coder-30B and prometheus-13b-v1.0 was $-0.675$ [95\% CI: $-0.723$, $-0.629$].

For Jensen-Shannon divergence, distributional differences were also almost universally significant after BH correction (adjusted p = 0.002 in nearly all comparisons), making this metric the most statistically decisive of the three metrics in separating model performance, with only isolated exceptions such as the DeepSeek-GRM-16B versus starcoder2-15b comparison in readability (p = 0.108) and complexity (p = 0.661). 

\begin{figure*}[t]
    \centering
  \includegraphics[width=0.85\linewidth]{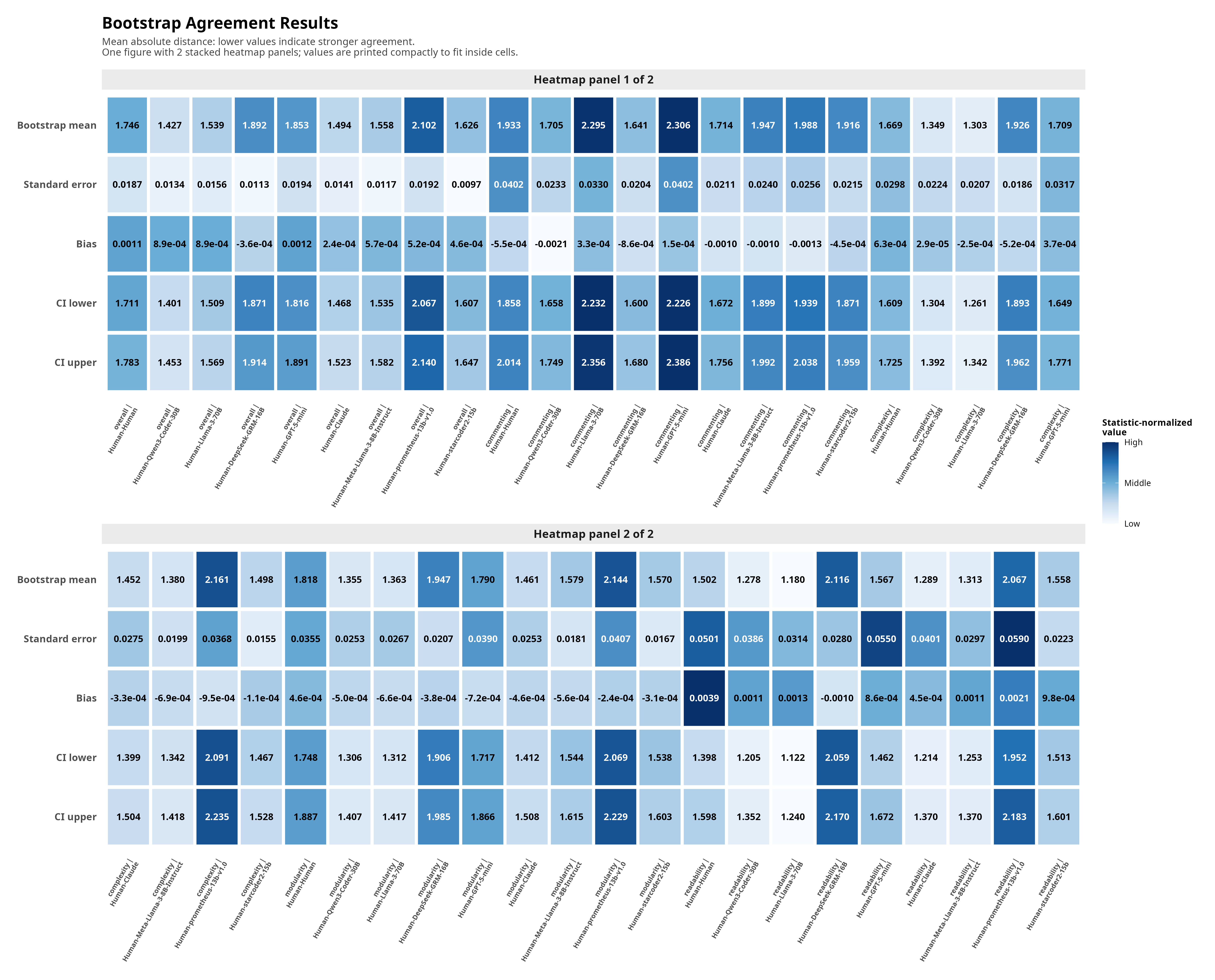}
  \caption {Human-LLM Bootstrapping Statistics for Disagreement metric.}
  \label{fig:bootstrap1}
\end{figure*}

\begin{figure*}[t]
    \centering
  \includegraphics[width=0.85\linewidth]{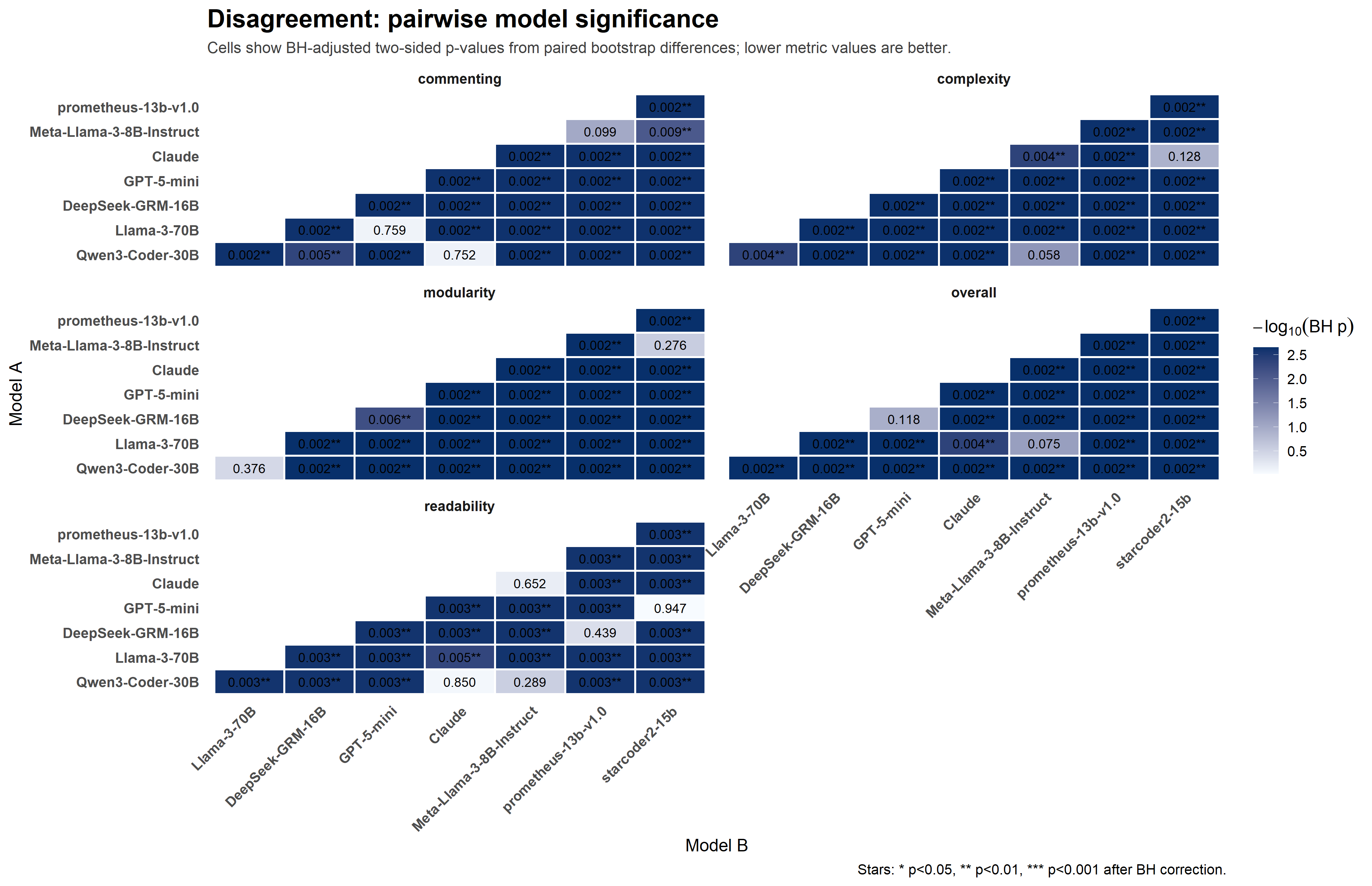}
  \caption {Pairwise models' significance scores for Disagreement metric.}
  \label{fig:significance1}
\end{figure*}

\begin{figure*}[t]
    \centering
  \includegraphics[width=0.85\linewidth]{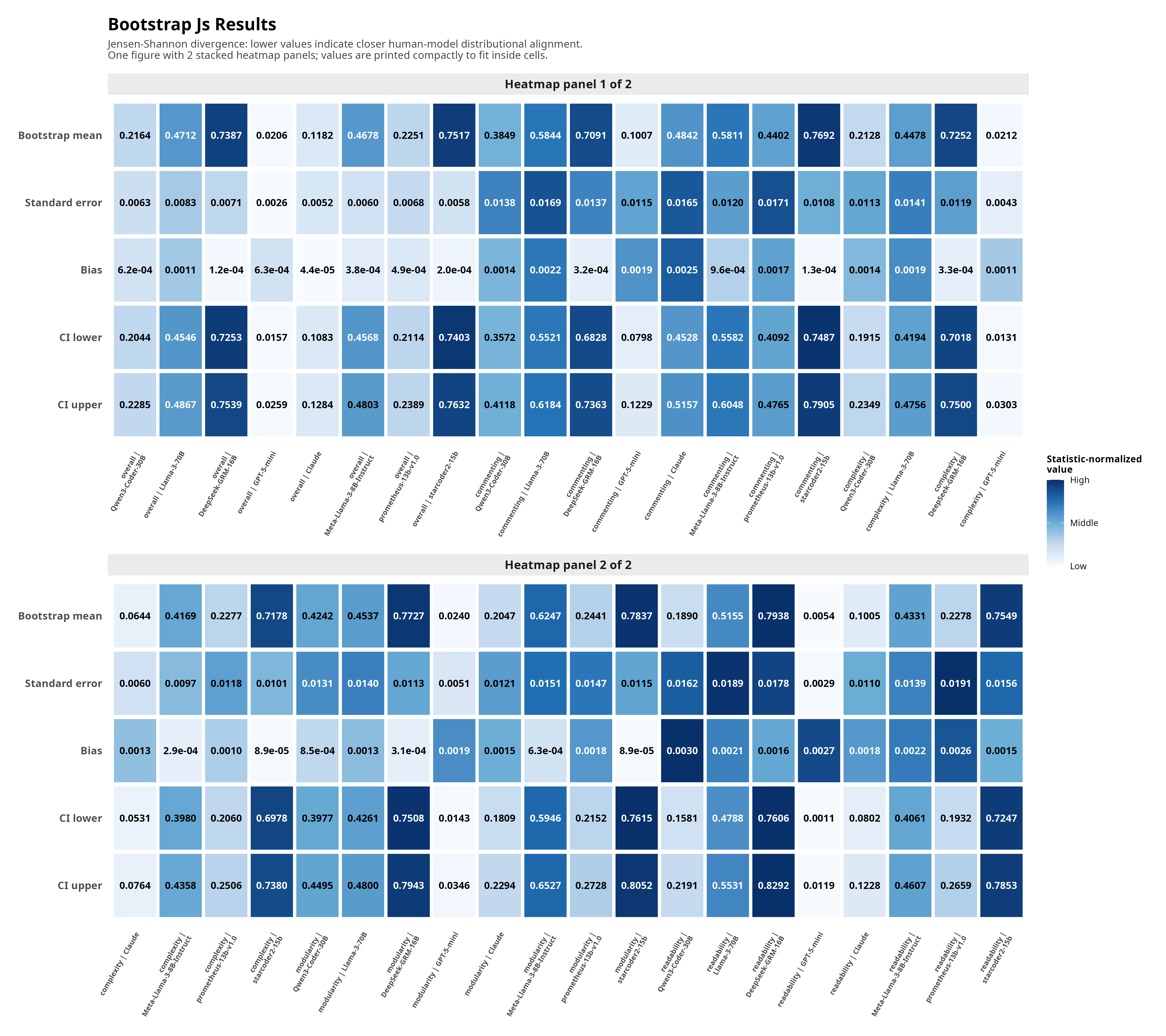}
  \caption {Human-LLM Bootstrapping Statistics for Jensen-Shannon divergence.}
  \label{fig:bootstrap2}
\end{figure*}

\begin{figure*}[t]
    \centering
  \includegraphics[width=0.85\linewidth]{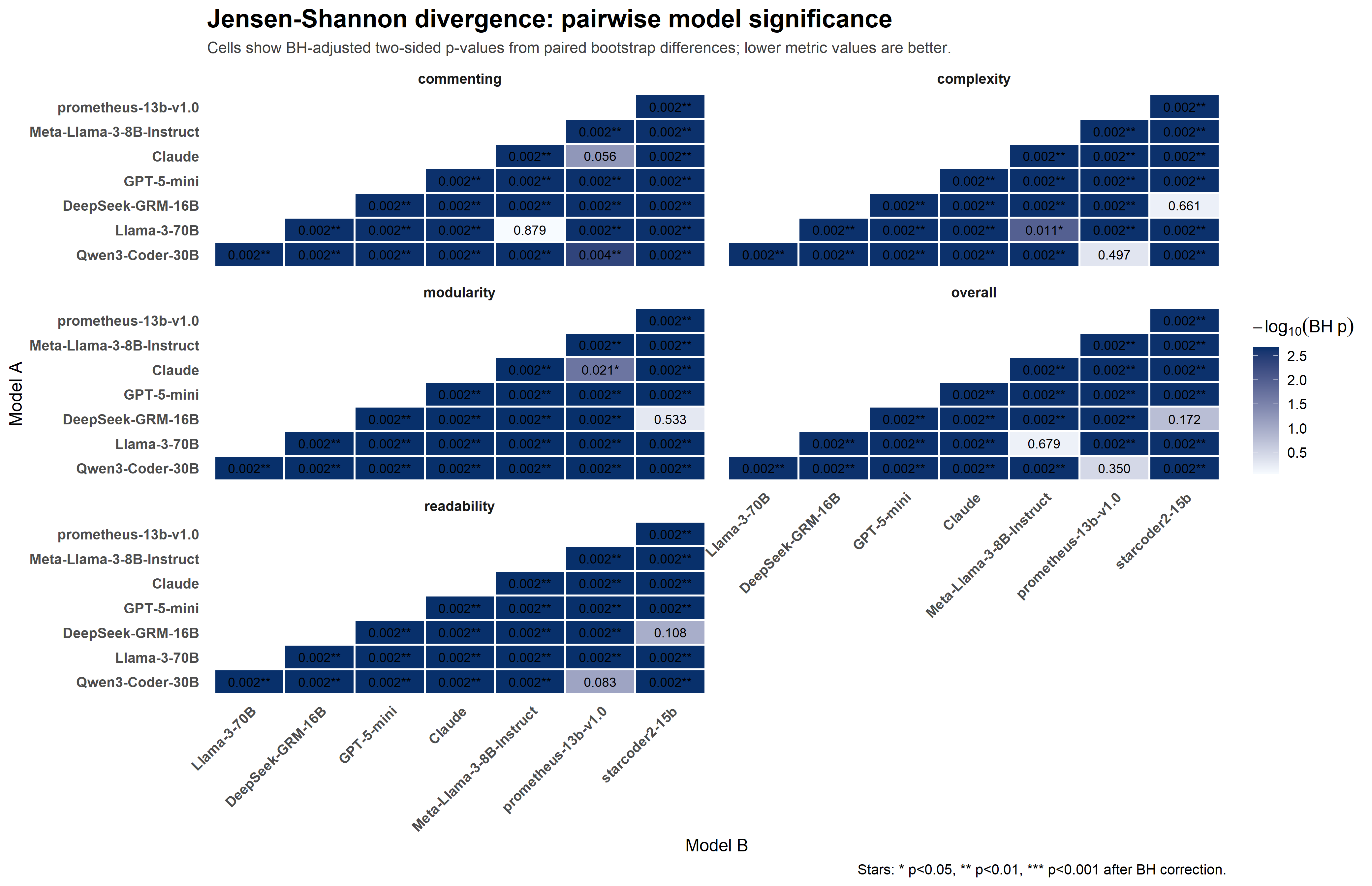}
  \caption {Pairwise models' significance scores for Jensen-Shannon divergence.}
  \label{fig:significance2}
\end{figure*}

\begin{figure*}[t]
    \centering
  \includegraphics[width=0.85\linewidth]{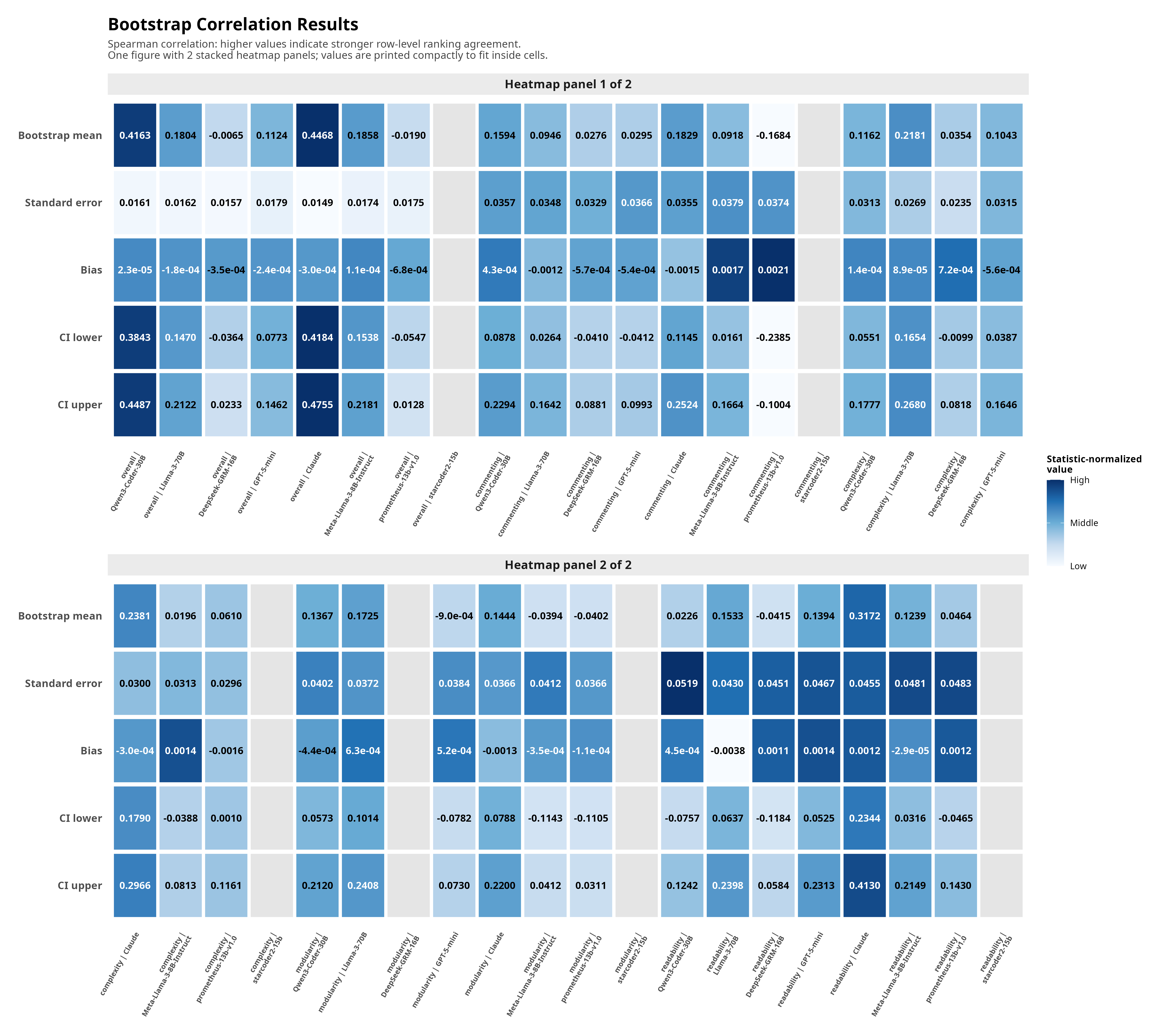}
  \caption {Human-LLM Bootstrapping Statistics for Spearman correlation.}
  \label{fig:bootstrap3}
\end{figure*}

\begin{figure*}[t]
    \centering
  \includegraphics[width=0.85\linewidth]{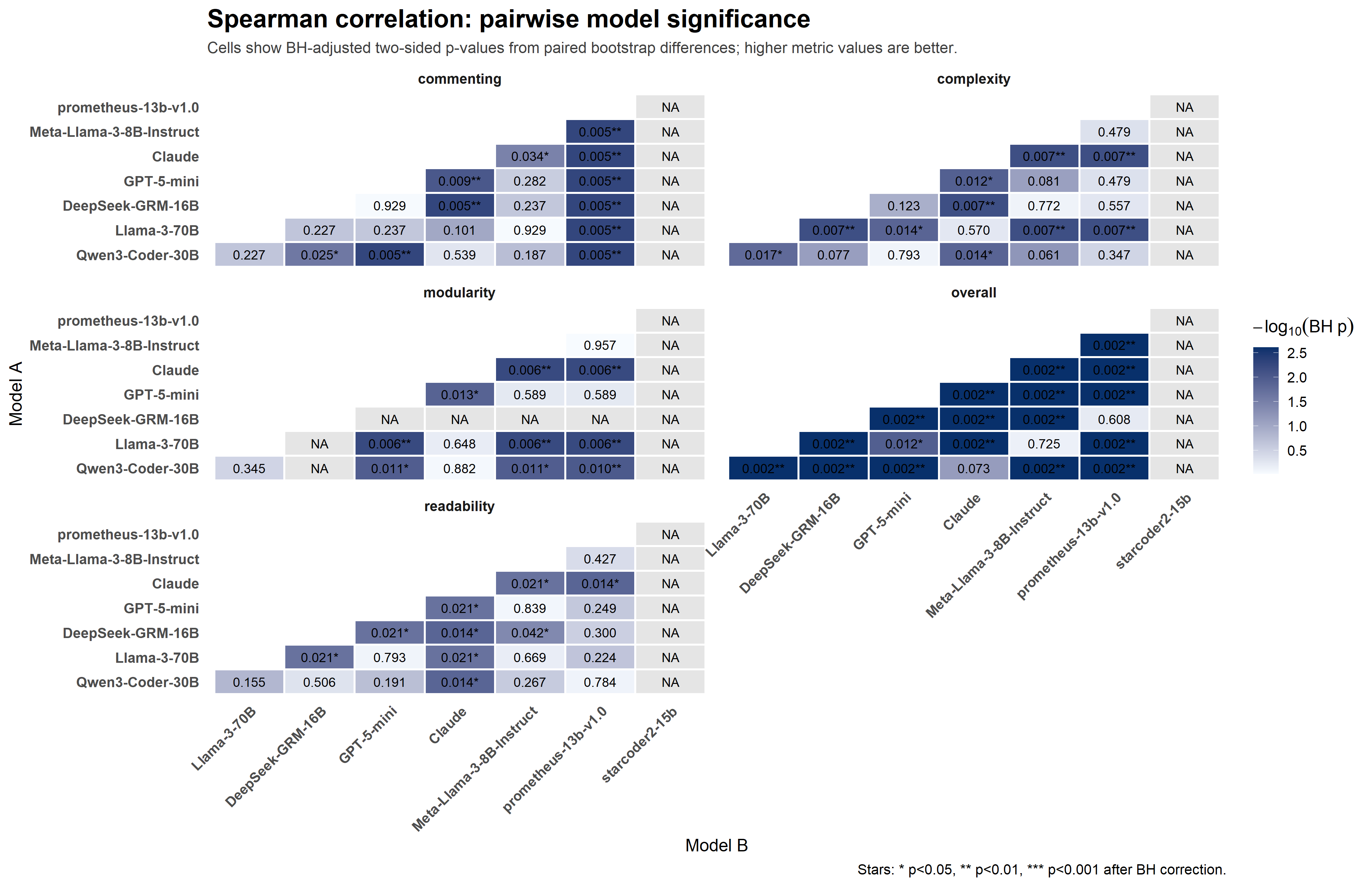}
  \caption {Pairwise models' significance scores for Spearman correlation.}
  \label{fig:significance3}
\end{figure*}

For Spearman correlation, Claude leads ($\rho = 0.447$, 95\% CI [0.418, 0.476], SE = 0.018), followed by QwenCoder ($\rho = 0.416$, 95\% CI [$0.384$, $0.449$], SE = 0.016); their difference is borderline significant (difference = $-0.031$, 95\% CI [$-0.062$, $-0.000$], p = 0.050), with Claude’s clearest advantage in readability ($\rho = 0.316$ vs. 0.022, difference = $-0.294$, 95\% CI [$-0.429$, $-0.162$], p = 0.010). For a fuller picture regarding the confidence intervals, biases, standard errors, and significance scores, refer to Figures \ref{fig:bootstrap1} to \ref{fig:significance3}.

\section{Examples for Qualitative analysis}\label{app:examples_qualitative_analysis}

\begin{enumerate}[label=\textbf{ID \arabic*.}]
\item
\textbf{Text\_A:} More computationally complex code\par
\textbf{Text\_B:} Less computationally complex code\par
\textbf{Annotator\_Score\_1:} 1\par
\textbf{Annotator\_Score\_2:} 3\par
\textbf{Annotator\_Score\_3:} 1\par
\textbf{GPT5\_Score:} 0\par
\textbf{Ann\_Comments\_1:} \mbox{}\par
\textbf{Ann\_Comments\_2:} I would choose Option B due to its greater efficiency and lower time complexity. While this function wouldn't most likely be used many times, by adding a large input on Option A the code would run recursively too many times, wasting time and resources.\par
\textbf{Ann\_Comments\_3:} \mbox{}\par
\textbf{GPT5\_Comments:} I picked 0 (strongly prefer A) because Code A is minimal, self-contained, and directly returns the gcd with very little implementation overhead, making it easy to read, verify, and use for typical small-to-moderate integers. Although the subtraction form can be asymptotically worse than the division-based/extended variant, Code B does extra work to produce Bezout coefficients and has more branching/sign handling (higher constant overhead), so unless you specifically need the coefficients or must handle very large numbers, the simpler implementation is preferable in the time/resources tradeoff for common use.\par
\textbf{Problem:} This Python code solves the algorithmic problem of GCD that involves computing the greatest common divisor of two integers using Euclid’s algorithm and returning the result.

\item
\textbf{Text\_A:} More computationally complex code\par
\textbf{Text\_B:} Less computationally complex code\par
\textbf{Annotator\_Score\_1:} 2\par
\textbf{Annotator\_Score\_2:} 3\par
\textbf{Annotator\_Score\_3:} 5\par
\textbf{GPT5\_Score:} 5\par
\textbf{Ann\_Comments\_1:} Even if more complex, the comments are much more precise. If the algorithm works, and especially is a known algorithm, then its okay to use it anywhere, as long as it is commented properly like in option B.\par
\textbf{Ann\_Comments\_2:} Option B shows the characteristics of dynamic programming: it uses overlapping subproblems and optimal substructures. It stores the dynamic programming parts in the dp array. It ends with a reconstructing step in the while loop where the string is again put together.\par
On the other hand, Option A does not reconstruct the string, nor is it dynamic programming. It is, as the first comment says, just a naive recursive implementation without any memoization. This makes it heavily computationally intense (exponentially).\par
However, option B is not perfect since it lacks online commenting that helps to explain the steps (overlapping subproblems, optimal substructures, reconstruction).\par
\textbf{Ann\_Comments\_3:} \mbox{}\par
\textbf{GPT5\_Comments:} I picked 5 (strongly prefer B) because the dynamic-programming implementation runs in O(mn) time and is practical for real inputs, whereas the naive recursive version is exponential and infeasible except for tiny strings. Code B also reconstructs the subsequence and handles edge cases cleanly (at the cost of O(mn) memory), making it far superior in the time/resources tradeoff.\par
\textbf{Problem:} This Python code solves the algorithmic problem of LCS that involves computing the longest common subsequence of two strings with dynamic programming and reconstructing the subsequence.

\item
\textbf{Text\_A:} More computationally complex code\par
\textbf{Text\_B:} Less computationally complex code\par
\textbf{Annotator\_Score\_1:} 0\par
\textbf{Annotator\_Score\_2:} 2\par
\textbf{Annotator\_Score\_3:} 2\par
\textbf{GPT5\_Score:} 0\par
\textbf{Ann\_Comments\_1:} \mbox{}\par
\textbf{Ann\_Comments\_2:} I would choose Option B because it implements Euclid’s algorithm in a much simpler and more readable recursive form while maintaining optimal time complexity. Option A is more verbose and harder to maintain, even though it computes additional coefficients (extended GCD). In daily work, clarity and simplicity are usually preferable when only the GCD value is required.\par
\textbf{Ann\_Comments\_3:} \mbox{}\par
\textbf{GPT5\_Comments:} I picked 0 (strongly prefer A) because Code A implements the plain Euclidean gcd with a single modulo recursion and minimal overhead, giving the fastest, simplest solution for just the GCD. Code B is an extended-GCD routine that maintains many extra variables and performs additional arithmetic to compute Bezout coefficients, so it does more work (and is more error-prone and harder to read) when only the gcd is required.\par
\textbf{Problem:} This Python code solves the algorithmic problem of GCD that involves computing the greatest common divisor of two integers using Euclid’s algorithm and returning the result.

\item
\textbf{Text\_A:} More computationally complex code\par
\textbf{Text\_B:} Less computationally complex code\par
\textbf{Annotator\_Score\_1:} 3\par
\textbf{Annotator\_Score\_2:} 2\par
\textbf{Annotator\_Score\_3:} 1\par
\textbf{GPT5\_Score:} 0\par
\textbf{Ann\_Comments\_1:} I would slightly prefer Option B. The dynamic programming approach efficiently computes the maximum value without generating all possible combinations, which makes it much faster for larger numbers of items. While Option A is conceptually simple, it quickly becomes impractical as the number of items grows, so Option B is more suitable for daily work where performance matters.\par
\textbf{Ann\_Comments\_2:} \mbox{}\par
\textbf{Ann\_Comments\_3:} \mbox{}\par
\textbf{GPT5\_Comments:} I chose 0 (strongly prefer A) because Code A directly returns both the optimal value and the explicit item selection (best\_combination), making it a complete, easy-to-verify solution, whereas the provided DP (Code B) only returns the maximum value and omits reconstruction of which items achieve it. Also, although brute-force is exponential, it is simple, uses built-in combinatorics (good for small n), and avoids the potentially large O(n·capacity) time/space cost and implementation pitfalls of the DP when capacity is huge or item reconstruction is required.\par
\textbf{Problem:} This Python code solves the algorithmic problem of Knapsack that involves selecting items to maximize value under a weight/capacity constraint using dynamic programming and returning both value and chosen items.

\item
\textbf{Text\_A:} More computationally complex code\par
\textbf{Text\_B:} Less computationally complex code\par
\textbf{Annotator\_Score\_1:} 1\par
\textbf{Annotator\_Score\_2:} 3\par
\textbf{Annotator\_Score\_3:} 1\par
\textbf{GPT5\_Score:} 0\par
\textbf{Ann\_Comments\_1:} \mbox{}\par
\textbf{Ann\_Comments\_2:} I would slightly prefer Option A. It uses a recursive merge sort approach, which has a predictable O(n log n) time complexity, making it more efficient for larger lists compared to bubble sort. Although the recursion and merging logic is a bit more complex, it’s generally more suitable for handling bigger datasets in daily work.\par
\textbf{Ann\_Comments\_3:} \mbox{}\par
\textbf{GPT5\_Comments:} I picked 0 (strongly prefer A) because this bubble\_sort is simple, in-place, and has low per-operation overhead with an early-exit optimization (best case O(n)), making it predictable and lightweight for typical small-to-moderate inputs. The provided merge\_sort implementation is actually inefficient in practice — it repeatedly slices lists, uses pop(0) (O(n) per pop), and concatenates lists, which induces extra copies and can degrade to much worse than the ideal O(n log n) behavior and use more memory, so its real-world time/space cost can exceed the bubble sort shown.\par
\textbf{Problem:} This Python code solves the algorithmic problem of Sorting Algorithm by ordering elements according to a key and stability rules, discussing time/space complexity and worst-case behavior.
\end{enumerate}

\end{document}